# Standing on FURM ground - A framework for evaluating Fair, Useful, and Reliable AI Models in healthcare systems

Alison Callahan[1,2], Duncan McElfresh[2], Juan M. Banda[2], Gabrielle Bunney[3], Danton Char[4,5], Jonathan Chen[1,6,7], Conor K. Corbin[1], Debadutta Dash[3], Norman L. Downing[1,6], Sneha S. Jain[8,9], Nikesh Kotecha[2,10], Jonathan Masterson[2], Michelle M. Mello[11,12,13], Keith Morse[14], Srikar Nallan[2], Abby Pandya[2], Anurang Revri[2], Aditya Sharma[2], Christopher Sharp[10], Rahul Thapa[2], Michael Wornow[15], Alaa Youssef[16], Michael A. Pfeffer[2,10], Nigam H. Shah[2,7,10]

[1]Center for Biomedical Informatics Research, Stanford University School of Medicine, Stanford, California, USA
[2]Technology and Digital Solutions, Stanford Health Care, Palo Alto, California, USA
[3]Department of Emergency Medicine, Stanford University School of Medicine, Stanford, California, USA
[4]Department of Anesthesia, Perioperative and Pain Medicine, Stanford University School of Medicine, Stanford, California, USA
[5]Stanford Center for Biomedical Ethics, Stanford University, Stanford, California, USA
[6]Division of Hospital Medicine, Stanford University School of Medicine, Stanford, California, USA
[7]Clinical Excellence Research Center, Stanford University School of Medicine, California, USA
[8]Division of Cardiovascular Medicine, Stanford University School of Medicine, Stanford, California, USA
[9]Cardiovascular Institute, Stanford University School of Medicine, Stanford, California, USA
[10]Department of Medicine, Stanford University School of Medicine, Stanford, California, USA
[11]Stanford Law School, Stanford University, Stanford, California, USA
[12]Department of Health Policy, Stanford University School of Medicine, Stanford, California, USA
[13]Freeman Spogli Institute for International Studies, Stanford University, Stanford, California, USA
[14]Department of Pediatrics, Stanford University School of Medicine, Stanford, California, USA
[15]Department of Computer Science, Stanford University, Stanford, California, USA
[16]Department of Radiology, Stanford University School of Medicine, Stanford, California, USA

# Abstract

The impact of using artificial intelligence (AI) to guide patient care or operational processes is an interplay of the AI model's output, the decision-making protocol based on that output, and the capacity of the stakeholders involved to take the necessary subsequent action. Estimating the effects of this interplay before deployment, and studying it in real time afterwards, are essential to bridge the chasm between AI model development and achievable benefit. To accomplish this, the Data Science team at Stanford Health Care has developed a Testing and Evaluation (T&E) mechanism to identify fair, useful and reliable AI models (FURM) by conducting an ethical review to identify potential value mismatches, simulations to estimate usefulness, financial projections to assess sustainability, as well as analyses to determine IT feasibility, design a deployment strategy, and recommend a prospective monitoring and evaluation plan. We report on FURM assessments done to evaluate six AI guided solutions for potential adoption, spanning clinical and operational settings, each with the potential to impact from several dozen to tens of thousands of patients each year. We describe the assessment process, summarize the six assessments, and share our framework to enable others to conduct similar assessments. Of the six solutions we assessed, two have moved into a planning and implementation phase. Our novel contributions – usefulness estimates by simulation, financial projections to quantify sustainability, and a process to do ethical assessments – as well as their underlying methods and open source tools, are available for other healthcare systems to conduct actionable evaluations of candidate AI solutions.

# Introduction

The impact of using artificial intelligence (AI) to guide clinical and operational workflows in healthcare systems is an interplay of an AI model's output, the decision-making protocol based on that output (e.g. intervene on a patient if their predicted risk of an outcome is > X%), the capacity of workflow participants to take action, and the benefits and harms of the action taken. The volume and mix of services provided, patient outcomes, and clinicians' work lives all may be affected by AI model deployment, with the potential for substantial ethical, financial, and workforce implications. Studying this interplay and its potential effects, both before and after implementation of a model-guided workflow, is essential for avoiding "pilotitis"[1].

Executive Order 14110 on AI[2,3], released by the Biden administration in October 2023, further motivates the need for tested, reproducible and rapid processes to assess the value, safety, and fairness of AI systems deployed in healthcare settings[4]. The executive order calls for standards to determine whether AI solutions and systems maximize benefit and minimize risk, and for policies on responsible use of AI in healthcare delivery[4]. This call has spurred researchers at academic medical centers, including Stanford, to propose a nationwide network of health AI assurance laboratories to develop a testing and evaluation (T&E) framework as well as corresponding processes and policies[5].

Existing strategies for testing and evaluating AI models in the context of the workflows they are part of consist primarily of reporting guidelines and checklists that focus on the properties of the model, including performance metrics, how it will be deployed, and potential outcomes[6,7]. There are limited T&E frameworks for assessing entire AI systems, beyond model performance, and minimal guidance on how to select the guidelines and checklists appropriate for a given situation. This requires the user to figure out how to combine the results of individual evaluations for decision-making—an unfilled gap commonly described as the "AI Chasm"[8]. Prior work to fill this gap has focused on developing evaluations for each step along the path from model development to deployment. At Stanford Medicine, the Responsible AI for Safe and Equitable Health (RAISE-Health) initiative launched in partnership with the Human Centered Artificial Intelligence Institute (HAI), guides the responsible use of AI across biomedical research, education, and patient care. A core part of the effort is establishing a framework for ethical health AI standards and safeguards.

Against this backdrop, the Data Science team in Technology and Digital Solutions at Stanford Medicine has developed a T&E mechanism — the <u>F</u>air, <u>U</u>seful, and <u>R</u>eliable AI <u>M</u>odel (FURM) assessment — to routinely estimate achievable benefits of AI model-guided workflows before deployment, which is necessary for bridging the AI Chasm. Every AI system proposed for deployment in our healthcare system requires a FURM assessment. This enables us to track AI project status, to be responsive to requests for information about the use of AI in our healthcare system (such from the California Attorney General in 2023), and to use FURM assessments to prioritize and triage proposed deployments.

In this article, we describe the FURM assessment process (Figure 1) and its use to evaluate six AI model use cases for potential deployment. While this process was designed for use at our institution, the novel components of this process — usefulness estimates by simulation, financial projections to quantify sustainability, and a process for conducting ethical assessments — are exportable to other healthcare systems, with open-source tools and methods that are feasible for other organizations to follow.

# Methods

## Overview of the FURM assessment process

We developed the FURM assessment to identify Fair, Useful and Reliable AI Models for workflows being considered for deployment in our healthcare system, Stanford Health Care (SHC). For each use case brought to the Data Science team, a FURM assessment is conducted in 3 stages: (1) **What & Why**, to assess potential usefulness, financial impact, and ethical considerations; (2) **How**, to identify requirements to deploy said AI model-guided workflow; and (3) **Impact**, to design experiments to monitor and evaluate the value of the AI model-guided workflow after deployment.

The 3-stage assessment process accommodates AI model-guided workflows at various stages of maturity: (1) not yet designed or deployed, (2) designed but not deployed, and (3) both designed and deployed. For those that are not yet designed or deployed, a Stage 1 assessment provides for early identification of AI model-guided workflows that are not aligned with an institutional or care need, do not have potential value in terms of financial or clinical outcomes, or have substantial ethical considerations that require further attention before deployment. For AI model-guided workflows that are designed but not yet deployed, Stage 2 and 3 assessments identify technical requirements for deployment, develop a monitoring plan, and outline evaluations to prospectively assess impact and ethical concerns. For AI-guided workflows that are already deployed, Stage 3 assessments provide a monitoring plan and outline for prospective evaluation. Going forward, all AI systems being considered for deployment will have a Stage 1 FURM assessment, followed by Stage 2 and 3 as appropriate. For deployed AI systems that predate the Data Science team, we can also conduct Stage 2 and/or Stage 3 assessments to inform modifications to deployment and impact evaluations.

The final product of each assessment is a report to support decisions of whether and how to proceed with deployment. Each report includes a recommendation for whether to proceed to the next stage and how each component of the assessment informs the recommendation. If we do not recommend proceeding, we discuss with the requesting team any modifications to the proposed AI model and workflow that would potentially change the recommendation, so that all recommendations are actionable. These assessments are shared with institutional leadership and the team that requested the assessment (see the Results section 'How we use FURM assessments'). SHC leadership uses FURM assessments to prioritize institutional support for potential AI deployments that have potential for positive impact and a sustainable financial profile.

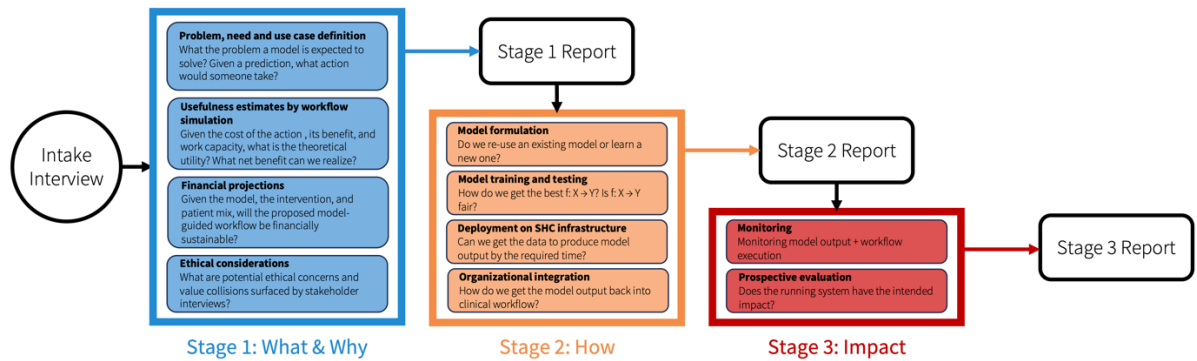

**Figure 1.** Stages and components of the FURM assessment. Each FURM assessment consists of 3 stages to evaluate the **what & why** motivating a particular AI use case, **how** a given AI model will be formulated, evaluated and integrated into a given healthcare system workflow, and the **impact** of the proposed implementation. Each of these 3 stages has multiple components (colored boxes), each addressing a different aspect of the use case.

**Participants in a FURM assessment.** Three sets of participants are involved in each assessment: *Requestor(s)*, an *Assessment Coordinator*, and *Component Leads*. The *Requestor(s)* are usually experts in the AI model and its motivating use case—such as clinicians who identified the use case, researchers who developed the AI model, or commercial vendors who have developed an AI-based product being considered for purchase by the healthcare system. The Data Science team also engages in needs finding efforts to engage with healthcare employees who are not experts in AI, but who do potentially face challenges or inefficiencies in their work that have the potential be improved with AI. The *Assessment Coordinator* is a member of the Data Science team who facilitates the assessment process, and is responsible for communication and logistical support, as well as compiling the final assessment report. *Component Leads* are subject matter experts responsible for each assessment component, and include clinicians, data scientists, IT professionals, financial analysts, ethicists, and healthcare administrators.

The FURM assessment process begins when the Requestor(s) bring a use case for an AI model to the Data Science Team. We first hold a "what to expect" session to orient the Requestor(s) to the process and what we will require from them to conduct an assessment. We then conduct an intake session with the Requestor(s), to obtain the information necessary to complete Stage 1 of our assessment. Following completion of the Stage 1 assessment, we make the decision of whether to recommend Stage 2 and 3 assessments, based on the potential impact and value of deployment as indicated by Stage 1.

The following sections describe each stage and component in detail; the Appendix includes templates of each component, as well as a completed report of each assessment stage for a single use case (with confidential information redacted). We describe the process as a linear series of components, however in our experience conducting six assessments to date, components may be completed simultaneously, sometimes in an iterative manner as we confirm assumptions or additional questions arise. We expect our FURM assessment process to continue to evolve, and so each component description and associated template may also change over time.

# Intake

Each assessment begins with an intake session, attended by the Requestor(s) and the assessment team. The purpose of this session is to establish rapport between the assessment team and the Requestor(s), and to set expectations for the FURM assessment process—including the responsibilities of each participant, the potential outcomes of the assessment, and the intended timeline to completion of the assessment.

# Stage 1: What and Why

Stage 1's main goal is to inform the decision of whether an AI model should be integrated into the proposed workflow and deployed. At this stage, an AI model does not yet have to be trained or implemented – it may only have been defined as a task for which a model will be trained to produce a specified outcome. Most use cases evaluated in Stage 1 involve an already existing model, but it is not a requirement. Three questions are answered at this stage: (1) *What problem does this AI model seek to solve?* (2) *How will the output of the model be used?* (3) *How might this model impact patients, personnel, and the healthcare system?* A Stage 1 FURM assessment consists of the following four components, designed to answer these questions: (1a) problem, need and use case definition, (1b) usefulness estimates by simulation, (1c) financial projections and (1d) ethical considerations. Component (1a) is completed first, because it provides essential information for the remaining components; (1b)-(1d) can be completed in any order.

**(1a) Problem, need and use case definition.** This component characterizes the purpose and intended use and impact of the AI model. This includes a description of the model-guided workflow (see Figure 2 for an example), the patient cohort, potential outcomes for patients, and the intended impact of the workflow on both patients and healthcare employees, including clinicians. This component also investigates whether there are existing alternatives to the AI model being proposed. A workflow diagram is created as part of this component to capture the decision points where the AI model output would be used to direct an action, including automating processes such as alerts or messages.

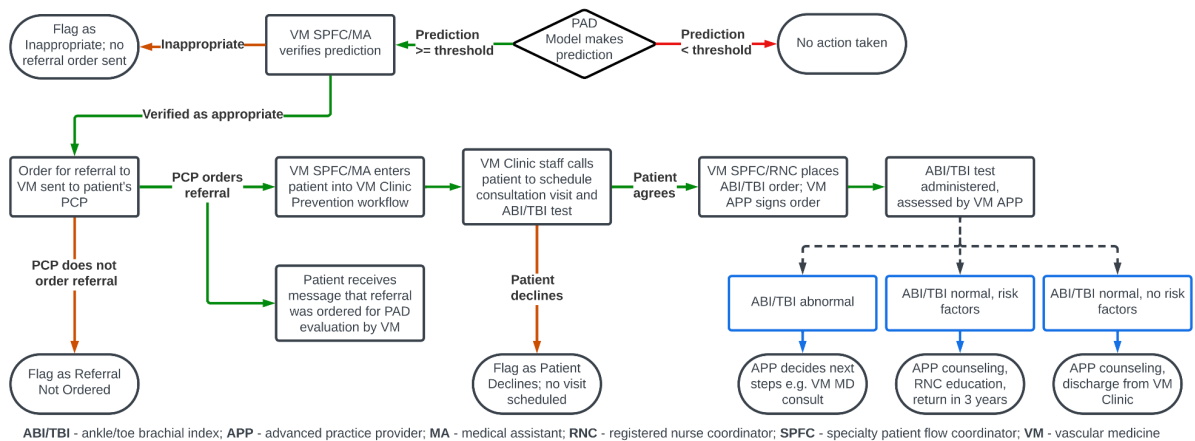

**Figure 2.** Example workflow diagram for component (1a): Problem, Need, and Use Case. In this example, a risk score for peripheral arterial disease (PAD) is used to identify patients for referral to Stanford Health Care's Vascular Medicine Clinic. Detailed workflows are essential for assessing the impact of an AI system on patients and clinicians.

**(1b) Usefulness estimates by workflow simulation.** This component uses a simulation tool, called APLUS[9], to generate quantitative estimates of the *achievable utility,* i.e. usefulness, of the proposed AI model-guided workflow. APLUS does so by considering the technical performance of the model alongside the capacity constraints of the team and setting where the model-guided workflow would be deployed, as well as the potential utility of the workflow. The utility of the workflow is highly dependent on the benefits and costs of possible outcomes[10–12], and thus the inputs to APLUS are specific to the use case. For example, for a use case aimed at reducing readmission, we might define utility in terms of the number of inpatient days that a patient experiences in the six months following initial discharge. For a use case focused on screening patients for ST-elevation myocardial infarction (STEMI) in the emergency department (ED), we would quantify utility based on patient wait time in the ED before being seen. We identify relevant utilities and capacity/resource constraints through the intake interview with the Requestor(s), use case-specific literature, and consultation with subject matter experts.

**(1c) Financial projections.** This component evaluates the potential financial impact of deploying and maintaining the proposed AI model-guided workflow. Importantly, this component characterizes the *differential* impact of deploying the AI model compared with existing state. Our analyses take into account the number of patients impacted, as well as the different possible outcomes of the AI model. Relevant costs include the cost of building (or purchasing) and maintaining the AI system, costs of personnel and resources to execute the model-guided workflow, and resulting costs of patient care or operational processes due to the model-guided workflow.

**(1d) Ethical considerations.** This component uses key informant interviews and ethical analysis to identify potential ethical issues relevant to the model itself *and* model-guided workflow impacts. In particular, this component identifies *value collisions* where key stakeholder groups may disagree on important model design or deployment characteristics[13,14]. We focus on potential adverse impacts on patient care and patients' interests, but also consider potential benefits and harms to patients, clinical staff, the healthcare system, and other stakeholders (e.g., a patient's family members, insurers). Distinctive ethical issues arise at different points in an AI model's life cycle, and following the "pipeline" framework[13], the ethical assessment process varies with the specific stage and nature of the AI model and associated workflow. To identify ethical considerations, we conduct semi-structured interviews with model developers, clinicians, patient representatives, and sometimes hospital administrators and IT staff. These interviews focus on 7 core principles: responsibility, equity, traceability, reliability, governance, nonmaleficence, and autonomy[15–17]. The interview guides continue to undergo refinement as we gain experience, but exemplar questions for each domain are provided in Table 1. Lastly, the ethics assessment team formulates, or requests that the development team formulate, potential AI model or workflow modifications to address specific ethical challenges identified. Ongoing work is considering how to iteratively consult the stakeholder representatives and a panel of AI experts to seek their feedback on the proposed interventions.

**Table 1.** The 7 core principles considered in the ethical considerations component of a Stage 1 FURM assessment.

| Principle | Corresponding example questions asked during FURM assessment |
| --- | --- |
| Responsibility | Who is taking responsibility for the performance of the AI model once deployed? Are there clear lines of accountability among the model developer, IT staff at the deploying organization, and clinical staff? How serious might the harm to the healthcare organization be if the model |
| | failed to perform as expected?[18] |
| Equity | Has the model been developed and tested using data that fairly represent the patient population at the deploying organization? Is there reason for concern that model performance may be better for some patient groups than others? Does the model have other potential equity implications (e.g., increasing healthcare costs for patients; disadvantaging particular patients or groups of patients in the healthcare system)?[19] |
| Traceability | Are the technologies and methods employed by the AI model transparent and auditable?[20] |
| Reliability | Has the AI model been tested for safety in settings similar to that being considered for deployment?[21] |
| Governance | Are there processes in place to detect or avoid harm, and is there a mechanism for humans to disengage or deactivate the AI model?[22] |
| Nonmaleficence | What evidence is there that the AI model would improve patient outcomes? What assumptions undergird the hoped-for benefits (especially regarding how clinicians will respond to the presence and output of the model), and are they realistic? In a worst-case scenario, might patient outcomes be harmed?  Is there any prospect of harm to patients' interests from sharing patient information with the model developer? |
| Autonomy | Does the AI model enhance or impinge on patients' decision making? What if anything should patients be told about use of the model in their care? |

# Stage 2: How

Stage 2 FURM assessments focus on the technical and organizational aspects of implementing an AI model-guided workflow. These considerations cannot be overlooked; infrastructure and organizational challenges can be a major barrier to implementing AI in healthcare settings[23,24]. To help practitioners overcome these barriers, Stage 2 assessments elicit information needed to implement the AI model and workflow in the setting appropriate to the use case. Findings from Stage 2 assessments help determine whether (and how) the AI model and workflow *can be deployed* in a healthcare system given available infrastructure and personnel. In some cases the answer to this question will be no—for example, if the required technical infrastructure does not exist. This stage consists of four components, outlined below.

**2(a) Model formulation.**  For most use cases evaluated with a FURM assessment, the relevant AI model already exists because it was developed and evaluated by the Requestor(s), or it is available from a vendor. However, for use cases assessed without an existing model, i.e. as a task for which a model will be trained, this component will recommend a model formulation that is appropriate for the use case e.g. predict the risk of a given outcome in a specified time frame, or classify patients as having a condition or not.

**(2b) Model training and testing.** This component will recommend methods for training and testing the model using data from the healthcare system, including a model learning algorithm e.g. Random Forest. Second, this component outlines an evaluation protocol for the AI model, and instructions for running this protocol on historical data. The evaluation protocol specifies metrics related to both model performance (such as positive predictive value or false negative rate) and clinical utility. The protocol also informs prospective evaluation and model monitoring, which are components of a Stage 3 FURM assessment (described below).

**(2c) Deployment on local infrastructure.** This component summarizes the processes and infrastructure necessary to deploy the AI model and workflow in the healthcare system. This includes information about the data sources (including the data's origin and update frequency), necessary data transformations, and how the AI model is to be deployed (e.g., whether it will run on premise in the healthcare system, in a cloud environment, or be accessed via a vendor's API). Automated alerts or messages triggered by the AI model workflow are described here as well. The Data Science team has designed archetypes consisting of platforms for model deployment and maintenance, data access, and pipelines for transferring model output back into the relevant compute environments and workflow systems (e.g., electronic health record, enterprise resource planning system), the most appropriate of which are recommended in this component. In completing this for use cases where an AI model exists and is available for use, we also determine whether the existing version can be deployed in a production environment or requires any refactoring (which would incur additional costs in terms of time and possibly financial resources).

**(2d) Organizational integration.** This component outlines the organizational aspects of deploying the AI model and workflow. First, this component identifies representatives of the healthcare system who will act as "owners" of the AI model and use case. This includes a *service line representative* to act as champion and point-of-contact for the model-guided workflow, a *budget owner* who will be responsible for financial decisions, and an *application owner* responsible for the computational infrastructure aspects of the proposed implementation. If the healthcare system requires a formal review of the AI model and workflow, this is described as well. Ideally, the service line representative and budget owner are identified during Stage 1 of the FURM assessment, and actively engaged in developing the integration strategy in Stage 2. Second, this component describes an implementation plan for the AI model and workflow[25], including outreach and training needed for successful adoption of the workflow by healthcare personnel[26] and a rollout plan. The rollout plan encompasses a strategy to align incentives with adoption and to define corresponding milestones.

# Stage 3: Impact

While Stages 1 and 2 address the potential impact and feasibility of an AI model-guided workflow prior to deployment, Stage 3 focuses on evaluating the *observed impact* during and after deployment. Ongoing monitoring is essential because the performance and effect(s) of AI models can change after deployment[27,28]. The two key components of this stage are a plan for prospective evaluation and a plan for monitoring of the AI model-guided workflow.

**(3a) Prospective evaluation.** In this component, prospective evaluation plans are developed for a "silent" deployment followed by a "loud" deployment using the healthcare system's infrastructure and personnel. In a silent deployment[29], an AI model is deployed on the infrastructure outlined in Stage 2 but *without* executing the workflow based on model output. For example, suppose that an AI model is to be applied for all patients admitted to the ICU, and an alert raised for patients that the model classifies as having sepsis. During silent deployment, the AI model will run for all patients admitted to the ICU, but no alerts or messages will be displayed to healthcare personnel. The silent deployment evaluation is intended answer the following questions:

- *Does the AI model process inputs and produce outputs as intended?*
- *Does the distribution of input data match expectations, when compared with retrospective analysis?*
- *Does the model output match expectations, when compared with retrospective analysis?* This step should use the evaluation protocol described in component (2c).
- *Does the distribution of input data or model output change over the course of the silent deployment?*

In a loud deployment, an AI model-guided workflow is implemented in its entirety, with healthcare personnel taking action in response to model output. This may be done in a randomized fashion for rigorous prospective evaluation. In addition to the questions listed above, the loud deployment evaluation plan is designed to assess impact on clinical and operational outcomes (including patient outcomes where applicable) and personnel behavior for a prespecified period of time. The outcomes measured in evaluating a loud deployment are identified in collaboration with the service line representative overseeing the AI model-guided workflow (who may be part of the requesting team, but not necessarily).

**(3b) Monitoring.** AI models must be monitored after deployment because their performance and impact are likely to change over time[28]. This component outlines a plan for monitoring the AI model and its workflow, including the following details:

- The required frequency of monitoring—e.g., each week, quarter, or year.
- A plan for monitoring *performance of the AI model*, including a description of all performance metrics and data required to calculate them. This should include the evaluation protocol described in component (2c).
- A plan for monitoring the *implementation of the AI-guided workflow*. This might include personnel adherence to the AI model-guided workflow, responses to model-generated alerts or messages, or other aspects of user interactions with the model flagged as raising ethical questions.
- A plan for monitoring the *impact of the AI model on key outcomes*, such as patient outcomes, treatment decisions, or productivity gains
- A plan for *updating or retiring the AI model*. This might recommend a schedule for updating model weights, or minimum-acceptable model performance thresholds to trigger model update via retraining[23].

## Debrief

After each of a Stage 1, Stage 2 and Stage 3 assessment, the Data Science team schedules a debrief session with the Requestor(s). During the debrief, we walk through each component of the FURM assessment report (see Appendix for a sample report), explaining how each component informs the overall recommendation and answering any questions that arise.

# Results

Having described the FURM assessment process, we now turn to key learnings to date. We begin by summarizing the outcomes of several completed FURM assessments. We then describe how FURM assessments are used in decision making at Stanford Health Care. We also describe the time required to complete a FURM assessment, how that informs the number of assessments that can be done simultaneously, and the total number that can be completed in a year.

## Completed FURM assessments

We have completed Stage 1 FURM assessments for six use cases, as well as Stage 2 and 3 assessments for one. The six use cases we have assessed span clinical (both inpatient and outpatient) and operational settings (Table 2), and have the potential to impact from approximately fifty up to tens of thousands patients each year. Two of the use cases (Screening for PAD, Improving Documentation and Coding for Inpatient Care) have moved into an implementation phase.

**Table 2.** Overview of six Stage 1 FURM assessments completed to date.

| Use case | Category | FURM assessment summary | | |
|---|---|---|---|---|
| | | Patients impacted per year | Financial projections | Ethical considerations (examples) |
| Screening for Peripheral Arterial Disease (PAD) | Clinical: outpatient | ~1400 | Independently sustainable | A stated goal is to decrease missed cases of PAD in non-white, non-male patients, but our data suggest the deployment population will be majority white. |
| Estimating Coronary Artery Calcium (CAC) from Computed Tomography (CT) Scans | Clinical: outpatient | ~5800 | Independently sustainable | There are potential issues of responsibility, e.g., who will manage the downstream use of CAC scores. The deployment population is largely white, and monitoring is needed to detect benefits/harms in minoritized groups. |
| Avoiding Unplanned Emergency Department (ED) Visits by Oncology Patients | Clinical: outpatient & inpatient | ~700 | Needs external support | Because the model was trained using data from patients who presented to the emergency department, the system may be biased to direct clinical resources to patients with a higher propensity to seek care, which could exacerbate health disparities. |
| Screening for ST-elevation Myocardial Infarction (STEMI) in the ED | Clinical: inpatient | ~150 | Needs external support | Will increasing the number of patients with STEMI identified without significantly increasing the number of ECGs being performed mean that other patients could have reduced access to ECG? |
| Improving Documentation and Coding for Inpatient Care | Operational | ~35 000 | Independently sustainable | The tool could result in increased billing, elevating patients' out-of-pocket costs. |
| Identifying Hospital-Acquired Sepsis | Clinical: inpatient | ~2 700 | Needs external support | The model has not been tested in a clinical setting, and furthermore it is unclear what "safety testing" means for this use case. |

# How FURM assessments are used

Completion of a Stage 1 FURM assessment triggers multiple follow-up actions. In addition to the Requestor(s) using the Stage 1 report in deciding whether to pursue implementation, the written Stage 1 report is shared with the Data Science Executive Committee, composed of SHC and School of Medicine leadership. The Executive Committee reviews Stage 1 reports in a dedicated monthly meeting, and recommends use cases for advancing to Stage 2 FURM assessments as well as for implementation using dedicated resources including computational infrastructure, technical support staff and financial support. Reports from Stage 2 assessments are similarly reviewed and used to recommend Stage 3 assessments and implementations. Note that not every use of AI advanced to implementation has to be independently sustainable to be recommended for advancement/implementation. Our goal is to identify additional costs that may be incurred and have the corresponding need for external support acknowledged and budgeted for.

Upon review by the Executive Committee, two of the six use cases we assessed have advanced to implementation: Screening for PAD and Improving Documentation and Coding for Inpatient Care. The Screening for STEMI use case was not recommended to advance because it is anticipated to impact a small number of patients each year, making the relative costs of AI implementation high. Similarly, the Avoiding Unplanned ED Visits by Oncology Patients use case has a high projected intervention cost due to the proposal to hire full time nursing staff to facilitate patient outreach as part of the intervention plan. The use case proposed to target patients with breast cancer, lung cancer, and lymphoma, which comprise a small proportion of oncology patients treated at SHC. The high intervention cost would thus be unsustainable given the relatively small number of patients projected to receive the outreach intervention, but this would change if, for example, patients with other types of cancer were included. For the Identifying Hospital-Acquired Sepsis use case, our financial projections suggest that the extreme tradeoff between the relative costs from false negatives (such as longer lengths of stay, based on the performance data available at the time of assessment) compared to the benefits realized from correctly identifying true positives would require external support to cover.

# AI governance

Healthcare system governance of AI systems (as distinct from similar efforts by developers and vendors) is essential for AI usefulness, efficiency and patient safety[30]. FURM assessments are one part of a multi-pronged effort by the Data Science team to bring effective AI solutions into clinical and operational workflows and to govern those processes. The common types of consultation requests the DS team receives span multiple product lifecycle stages:

- Uncovering and understanding enterprise pain points to prioritize for FURM assessments. We engage in needs finding by interviewing business and clinical leaders, conducting brainstorming sessions at operational team meetings, holding targeted discovery meetings and shadowing healthcare system employees as they do their work (in patient care settings and others).
- Performing diligence on the business need and/or technical feasibility validation of AI solutions and deciding whether to build or buy.
- Conducting FURM assessments to evaluate potential AI projects.
- Operationalizing using our infrastructure to build and deploy model-guided workflows.

- Monitoring and evaluating deployed systems to analyze trends and investigate performance, impact and value over time.

These efforts in turn feed into projects overseen by the Stanford Medicine Technology and Digital Solutions (TDS) team. TDS manages all information technology applications at SHC, including maintaining a record of all deployed AI systems, securing approval for new projects for deployment, and deciding how to allocate TDS resources to such projects. FURM assessments are now used as part of TDS governance processes in making these approval and resource allocation decisions.

## Time required to complete a FURM assessment

With an overall goal to see 3-5 projects through to deployment per year, the Data Science team is currently staffed to complete one FURM assessment (Stage 1, Stage 2 or Stage 3) per month. Using Little's Law (a theorem used to determine the average number of items in a stationary queueing system[31]), and the approximate amount of time required for each component of a Stage 1 assessment (Figure 3), we estimate that this throughput requires conducting 2 assessments simultaneously at any given time.

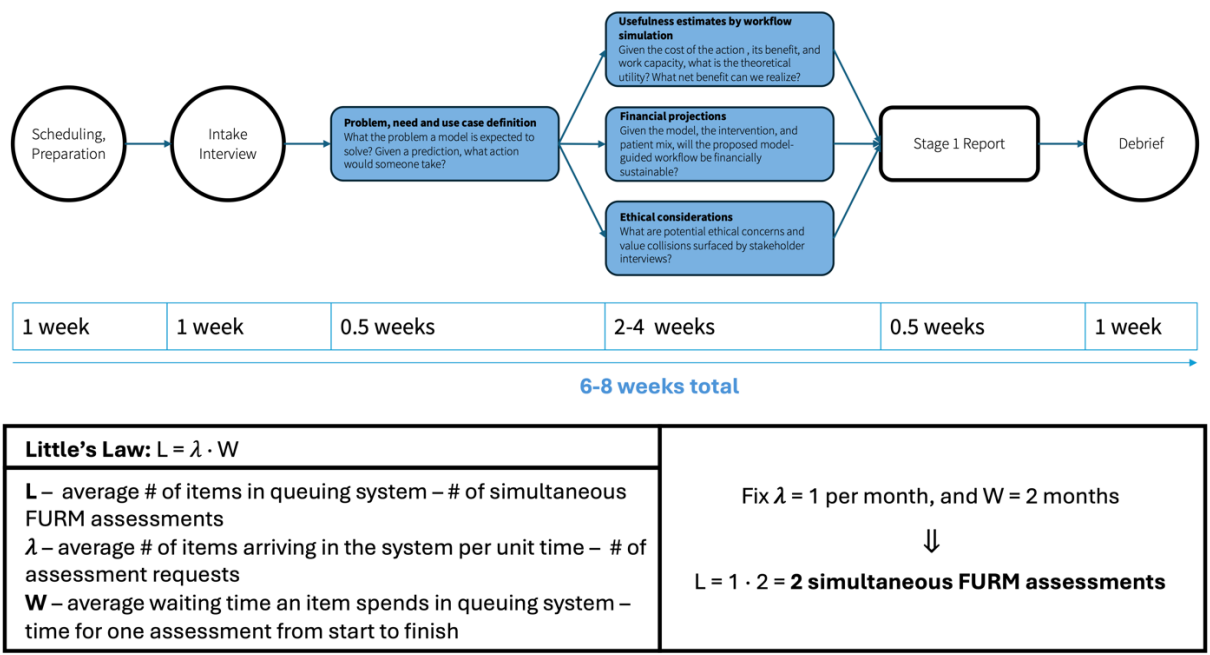

**Figure 3.** Estimated time required to complete each component of a Stage 1 FURM assessment, and application of Little's Law to arrive at a desired cadence of 2 simultaneous FURM assessments, in order to complete one FURM assessment per month, with the ultimate goal of seeing 3-5 potential AI guided workflows through to deployment in the healthcare system.

# Discussion

Direct involvement of healthcare system leadership in decisions to deploy AI systems is essential to use AI responsibly, avoid patient harm, and facilitate the long-term sustainability of AI implementations. Effective decision-making by leadership requires a rapid, repeatable and flexible assessment process whose results can be distilled into an executive summary for leadership, clinical and technical audiences. FURM assessments play this role at our healthcare system. Top-down support both of AI models and of their pre- and post-deployment evaluation ensures that the AI solutions healthcare systems invest in are sustainable, and have strong potential to improve patient care and/or operational endpoints. Post-deployment evaluation also helps to inform the equally important decision of when an AI model should be decommissioned or modified.

FURM assessments synthesize and build upon a foundation of existing guidance around T&E frameworks. These include the Key Decision Points developed by the Health AI partnership[32,33], the Coalition for Health AI's Blueprint for Trustworthy AI Implementation Guidance and Assurance for Healthcare[34], and the Duke Algorithm-Based Clinical Decision Support (ABCDS) Oversight framework[35]. The existing frameworks provide guidance for identifying and prioritizing use cases, evaluating AI models for performance and safety, designing workflows, implementing the proposed system, as well as monitoring and outcome measurement post-deployment. We contribute three novel elements to existing testing and evaluation regimens: pre-deployment ethical review for identification of stakeholder value mismatches, financial sustainability analysis, and the use of simulation to estimate achievable benefit under a given work capacity[9].

With an estimated cost of over $200,000 to integrate a single AI model into a workflow[36], such pre-deployment analyses are essential. We anticipate that the assessment processes we have developed can be adopted into existing frameworks such those by HAIP, CHAI and the National Academy of Medicine's effort to develop an AI Code of Conduct. By freely sharing the assessment process, underlying methods[12] and open source tools[9] used to quantify realizable benefits, we make it possible for other healthcare systems to conduct actionable evaluations of candidate AI systems.

The FURM assessment process itself evolved over the first six assessments, including updates to the software as well as procedural components to increase efficiency and consistency. For example, we added the structured What To Expect and Debrief sessions (before and after, respectively) to the assessment, analogous to a pre-visit survey and post-visit instructions that occur with clinical visits. We updated the APLUS software to support additional use cases, and revised the report layout based on feedback from requesting teams. Work to further expand the ethics assessment component continues.

As described above, our executive leadership uses FURM assessments as part of their decision-making to prioritize projects. Of the six assessment reports that have been reviewed, the emerging pattern of what is deemed a "worthy project" are situations where a large number of patients are expected to benefit, the implementation is projected to be financially sustainable (not necessarily having a positive margin), and the ethical concerns are such that a monitoring plan can be devised to measure their impact or be remedied via workflow modifications.

The FURM assessment process has several limitations. First, we have conducted FURM assessments for a limited number of use cases and adjustments may be needed for other kinds of use cases. For example, the Data Science team was asked to assess a use case involving the use of generative AI (specifically GPT-4) for drafting replies to messages patients send to clinicians through the healthcare system's online patient portal. A usefulness estimate and financial projections were not needed, but identifying ethical considerations was paramount. Thus, we conducted a "partial" FURM assessment focused on ethical considerations so the project could proceed to a pilot phase. Second, FURM financial projections require access to financial data. We work directly with members of the SHC Finance Department to identify and analyze financial data specific to each use case we assess. Such cross-departmental collaboration is necessary for successfully implementing the FURM assessment process, but may be more difficult to achieve at other institutions. Finally, only one component of our Stage 1 FURM assessment process (usefulness estimates by simulation) is available as software. The other components rely on interviews, qualitative evaluation, and custom financial analyses, practices for which can only be shared with other institutions as playbooks and interview guides. As members of the Coalition for Health AI, we aim to disseminate our approach through playbooks and interview guides, and to offer our services as part of the recently proposed nationwide network of AI assurance labs[5].

# Conclusion

We have developed a process to identify fair, useful and reliable AI models (FURM) by first concretizing the problem to be solved by a given AI model-guided workflow, and then conducting simulations to estimate usefulness, financial projections to evaluate sustainability, and expert interviews to identify potential stakeholder value mismatches and relevant ethical considerations. We also assess each proposed AI model-guided workflow to determine implementation feasibility, design a deployment strategy, and develop a prospective monitoring and impact evaluation plan. We have conducted FURM assessments for six use cases, and recommended two to advance to deployment. Our novel contributions – usefulness estimates by simulation, financial projections to quantify sustainability, and ethical considerations – as well as their underlying methods[12] and open source tools[9], make it possible for other healthcare systems to conduct actionable evaluations of candidate AI solutions.

# References

1. Huang F, Blaschke S, Lucas H. Beyond pilotitis: taking digital health interventions to the national level in China and Uganda. Global Health 2017;13(1):49.

2. The White House. Executive order on the safe, secure, and trustworthy development and use of artificial intelligence [Internet]. The White House. 2023 [cited 2024 Jan 29];Available from: https://www.whitehouse.gov/briefing-room/presidential-actions/2023/10/30/executive-order-on-the-safe-secure-and-trustworthy-development-and-use-of-artificial-intelligence/

3. The White House. Remarks by president Biden and vice president Harris on the administration's commitment to advancing the safe, secure, and trustworthy development and use of artificial intelligence [Internet]. The White House. 2023 [cited 2024 Jan 29];Available from: https://www.whitehouse.gov/briefing-room/speeches-remarks/2023/10/30/remarks-by-president-biden-and-vice-president-harris-on-the-administrations-commitment-to-advancing-the-safe-secure-and-trustworthy-development-and-use-of-artificial-intelligence/

4. Mello MM, Shah NH, Char DS. President Biden's executive order on artificial intelligence-implications for health care organizations. JAMA [Internet] 2023;Available from: http://dx.doi.org/10.1001/jama.2023.25051

5.  Shah NH, Halamka JD, Saria S, et al. A Nationwide Network of Health AI Assurance Laboratories. Journal of the American medical 2024;331(3):245–9.

6.  Lu JH, Callahan A, Patel BS, et al. Assessment of adherence to reporting guidelines by commonly used clinical prediction models from a single vendor. JAMA Netw Open 2022;5(8):e2227779.

7.  de Hond AAH, Leeuwenberg AM, Hooft L, et al. Guidelines and quality criteria for artificial intelligence-based prediction models in healthcare: a scoping review. NPJ Digit Med [Internet] 2022;5(1). Available from: http://dx.doi.org/10.1038/s41746-021-00549-7

8.  Marwaha JS, Kvedar JC. Crossing the chasm from model performance to clinical impact: the need to improve implementation and evaluation of AI. NPJ Digit Med 2022;5(1):25.

9.  Wornow M, Gyang Ross E, Callahan A, Shah NH. APLUS: A Python library for usefulness simulations of machine learning models in healthcare. J Biomed Inform 2023;139:104319.

10. Baker SG, Cook NR, Vickers A, Kramer BS. Using relative utility curves to evaluate risk prediction. J R Stat Soc Ser A Stat Soc 2009;172(4):729–48.

11. Ko M, Chen E, Agrawal A, et al. Improving hospital readmission prediction using individualized utility analysis. J Biomed Inform 2021;119(103826):103826.

12. Singh K, Shah NH, Vickers AJ. Assessing the net benefit of machine learning models in the presence of resource constraints. J Am Med Inform Assoc 2023;30(4):668–73.

13. Char DS, Abràmoff MD, Feudtner C. Identifying ethical considerations for machine learning healthcare applications. Am J Bioeth 2020;20(11):7–17.

14. Cagliero D, Deuitch N, Shah N, Feudtner C, Char D. A framework to identify ethical concerns with ML-guided care workflows: a case study of mortality prediction to guide advance care planning. J Am Med Inform Assoc 2023;30(5):819–27.

15. Defense Innovation Board. AI Principles: Recommendations on the ethical use of artificial intelligence by the Department of Defense. Department of Defense; 2019.

16. Abràmoff MD, Tobey D, Char DS. Lessons learned about autonomous AI: Finding a safe, efficacious, and ethical path through the development process. Am J Ophthalmol 2020;214:134–42.

17. Abràmoff MD, Cunningham B, Patel B, et al. Foundational considerations for artificial intelligence using ophthalmic images. Ophthalmology 2022;129(2):e14–32.

18. Price WN II, Gerke S, Cohen IG. Potential liability for physicians using artificial intelligence. JAMA 2019;322(18):1765.

19. Rajkomar A, Hardt M, Howell MD, Corrado G, Chin MH. Ensuring fairness in machine learning to advance health equity. Ann Intern Med 2018;169(12):866.

20. Funer F. The deception of certainty: How non-interpretable Machine Learning outcomes challenge the epistemic authority of physicians. A deliberative-relational approach. Med Health Care Philos 2022;25(2):167–78.

21. Challen R, Denny J, Pitt M, Gompels L, Edwards T, Tsaneva-Atanasova K. Artificial intelligence, bias and clinical safety. BMJ Qual Saf 2019;28(3):231–7.

22. Vayena E, Blasimme A. Health research with big data: Time for systemic oversight. J Law Med Ethics 2018;46(1):119–29.

23. Watson J, Hutyra CA, Clancy SM, et al. Overcoming barriers to the adoption and implementation of predictive modeling and machine learning in clinical care: what can we learn from US academic medical centers? JAMIA Open 2020;3(2):167–72.

24. D'Hondt E, Ashby TJ, Chakroun I, Koninckx T, Wuyts R. Identifying and evaluating barriers for the implementation of machine learning in the intensive care unit. Commun Med (Lond) [Internet] 2022;2(1). Available from: http://dx.doi.org/10.1038/s43856-022-00225-1

25. Bauer MS, Kirchner J. Implementation science: What is it and why should I care? Psychiatry Res 2020;283(112376):112376.

26. Leeman J, Birken SA, Powell BJ, Rohweder C, Shea CM. Beyond "implementation strategies": classifying the full range of strategies used in implementation science and practice. Implement Sci [Internet] 2017;12(1). Available from: http://dx.doi.org/10.1186/s13012-017-0657-x

27. Davis SE, Lasko TA, Chen G, Siew ED, Matheny ME. Calibration drift in regression and machine learning models for acute kidney injury. J Am Med Inform Assoc 2017;24(6):1052–61.

28. Davis SE, Walsh CG, Matheny ME. Open questions and research gaps for monitoring and updating AI-enabled tools in clinical settings. Front Digit Health 2022;4:958284.

29. Tonekaboni S, Morgenshtern G, Assadi A, et al. How to validate Machine Learning Models Prior to Deployment: Silent trial protocol for evaluation of real-time models at ICU. In: Flores G, Chen GH, Pollard T, Ho JC, Naumann T, editors. Proceedings of the Conference on Health, Inference, and Learning. PMLR; 07--08 Apr 2022. p. 169--82.

30. Nong P, Hamasha R, Singh K, Adler-Milstein J, Platt J. How academic medical centers govern AI prediction tools in the context of uncertainty and evolving regulation. NEJM AI [Internet] 2024;Available from: http://dx.doi.org/10.1056/aip2300048

31. Dongarra J, Luszczek P, Feautrier P, et al. Little's Law. In: Encyclopedia of Parallel Computing. Boston, MA: Springer US; 2011. p. 1038–41.

32. Key decision points [Internet]. Health AI Partnership. 2023 [cited 2024 Jan 29];Available from: https://healthaipartnership.org/key-decisions-in-adopting-an-ai-solution

33. Kim JY, Boag W, Gulamali F, et al. Organizational governance of emerging technologies: AI adoption in healthcare [Internet]. arXiv [cs.AI]. 2023 [cited 2024 Jan 29];Available from: http://arxiv.org/abs/2304.13081

34. Coalition for Health AI. Blueprint for Trustworthy AI Implementation Guidance and Assurance for Healthcare [Internet]. 2023;Available from: https://www.coalitionforhealthai.org/papers/blueprint-for-trustworthy-ai_V1.0.pdf

35. Bedoya AD, Economou-Zavlanos NJ, Goldstein BA, et al. A framework for the oversight and local deployment of safe and high-quality prediction models. J Am Med Inform Assoc 2022;29(9):1631–6.

36. Sendak MP, Balu S, Schulman KA. Barriers to achieving economies of scale in analysis of EHR data. A cautionary tale. Appl Clin Inform 2017;8(3):826–31.

# Appendix

FURM Assessment Report Templates

**Stanford** | **MEDICINE** | School of Medicine

| FURM Assessment Stage 1 Summary: [NAME OF PROJECT] - Stanford Health Care |
|---|

**Problem:** [ADD]

**Need:** [ADD]

**Eligible patient population:** [ADD]

**Proposed Model and Workflow Details**

**Model Specification**
- Outcome: [ADD]
- Output: [ADD]
- Training data source: [ADD]
- Deployment data source: [ADD]

**Intervention**
- 

**Workflow Constraints**
- Personnel capacity
  - 
- Time requirements
  - 

**Financial Projections**

| Y0 (Deployment) | Y1 | Y2 | Y3 | Y4 | Y5 |
|---|---|---|---|---|---|
| $ | $ | $ | $ | $ | $ |

- Revenue: [ADD DETAIL ON MAJOR REVENUE DRIVERS]
- Costs: [ADD DETAIL ON MAJOR COST DRIVERS]
- Sensitivity: [ADD DETAIL ON PROJECTION SENSITIVITIES]
- Of note: [ADDITIONAL RELEVANT CONTEXT]

**Potential Impact for Patients or Providers**
[ADD]

**Usefulness Estimates by Workflow Simulation**
[ADD]

**Ethical Considerations**
[ADD]

**Recommendation: [OVERALL RECOMMENDATION]**
- [ADD CONTEXT FOR RECOMMENDATION]

**Next Steps**
- [ADD]

**References**
**[ADD USE CASE SPECIFIC REFERENCES]**

**Evaluating AI/ML deployments**

Morse K, Bagley SC, Shah NH. Estimate the hidden deployment cost of predictive models to improve patient care. Nat Med 2020 Jan;26(1):18-19.

**Ethics for AI/ML deployments**

Char D, Shah N, Magnus D. Implementing Machine Learning in Health Care — Addressing Ethical Challenges. New Eng J Med, 2018 Mar 15;378(11):981-983.

Cagliero D, Deuitch N, Shah N, Char D. Evaluating Ethical Concerns with Machine Learning to Guide Advance Care Planning. Journal of Investigative Medicine, 2021;69:152

**Authors and Contributors (alphabetical by last name)**

# Problem, Need and Use Case Details

## Problem Statement
[ADD]

## Needs Statement
[ADD]

## Eligible Patient Population
[ADD]

## Proposed Intervention
[ADD]

## Intended Impacts on Patients
- [ADD]

## Potential Unintended Impacts on Patients and Providers
- [ADD]

## Proposed Workflow
[ADD, WITH FIGURE]

### Workflow Assumptions
- Capacity of personnel (e.g. calls a nurse can make per day)
  - 
- Time requirements for relevant activities
  - 

## Utilities of Outcomes
- **False positive**:
- **False negative**:
- **True positive**:
- **True negative**:

# Usefulness assessments via simulation

## Overview
[ADD]

## Utility Estimates

## Usefulness Simulations

## Summary
[ADD]

## References
[ADD]

Stanford **MEDICINE** | School of Medicine

# Financial Projections

## Overall Assessment

| Y0 (Deployment) | Y1 | Y2 | Y3 | Y4 | Y5 |
|---|---|---|---|---|---|
| $ | $ | $ | $ | $ | $ |

Recommendation – **[OVERALL RECOMMENDATION]**
- [ADD CONTEXT FOR RECOMMENDATION]

## Assumptions

**Cohort**
- 

**Model & Workflow**
- 

**Volume**
- 

**Financial**
- 

## Sensitivities

| Most Sensitive to 10% increase in: | Least Sensitive to 10% increase in: |
|---|---|
|  |  |
|  |  |
|  |  |
|  |  |

Stanford MEDICINE | School of Medicine

# Ethical Considerations

We assess the proposed intervention according to 9 ethical principles: *Responsibility*, *Equity*, *Traceability*, *Reliability*, *Governance*, *Non-Maleficence*, and *Autonomy*.

## Responsibility

*Humans should exercise judgment & remain responsible for use & outcomes.*

[ADD]

## Equity

*Avoid unintended bias & inadvertent harm; fairness in distribution, access and benefits across patient groups.*

[ADD]

## Traceability

*Transparent & Auditable methodologies, data sources, design procedures.*

[ADD]

## Reliability

*Explicit domain of use; safety tested across the entire life cycle of use in that domain.*

[ADD]

## Governance

*Possess the ability to detect/avoid unintended harm & for human disengagement or deactivation.*

[ADD]

## Non-Maleficence

*Do no harm; patient benefit; improved clinical outcomes.*

[ADD]

## Autonomy

*Patient still in control of their healthcare; liability for AI system malfunction related to degree of autonomy; ownership of data.*

**Quality of the Source:** [ADD]

**Quality of the Information:** [ADD]

## References

1. Fenton, Joshua J., et al. "Influence of computer-aided detection on performance of screening mammography." *New England Journal of Medicine* 356.14 (2007): 1399-1409.

| **FURM Assessment Stage 2 Summary: [PROJECT NAME] - Stanford Health Care** | |
|---|---|
| **Problem:** [ADD]<br><br>**Need:** [ADD]<br><br>**Eligible patient population:** [ADD] | |
| **Model formulation, training and testing** | |
| **Task**<br>•<br>**Model learning algorithm**<br>•<br>**Data**<br>•<br>**Testing and training splits**<br>• | **Cross Validation**<br>•<br>**Performance metrics**<br>•<br>**Fine-tuning procedures**<br>•<br>**Model validation**<br>• |
| **Deployment on SHC infrastructure** | |
| **Organizational integration** | |
| **Recommendation: [OVERALL RECOMMENDATION]**<br>• [ADD CONTEXT FOR RECOMMENDATION] | |
| **Next Steps**<br>• [ADD] | |
| **References**<br>[ADD] | |
| **Authors and Contributors (alphabetical by last name)**<br>[ADD] | |

# Model formulation

[ADD]

**Data labeling criteria:**

**Cohort entry criteria**: [ADD]

**Cases**: [ADD]

**Controls**: [ADD]

**General exclusions:** [ADD]

# Model training and testing

## Data variables

[ADD]

## Testing and training splits

[ADD]

## Cross Validation details

[ADD]

## Performance metrics

[ADD]

## Fine-tuning procedures

[ADD]

## Model validation

[ADD]

## Deployment on SHC infrastructure

[ADD]

# Organizational integration

[ADD]

## Clinical and operational integration

[ADD]

## People roles and responsibilities

| Role | Responsibilities | Name, Title |
|------|------------------|-------------|
| Service Line Owner | Budget owner who will be responsible for financial decisions | |
| Physician Champion | Leader and point-of-contact for the model-guided workflow | |
| TDS Champion | Application owner responsible for IT aspects of the proposed implementation | |

## Workflow

[ADD]

| FURM Assessment Stage 3 Summary: [PROJECT NAME] - Stanford Health Care |
|---|

**Problem:** [ADD]

**Need:** [ADD]

**Eligible patient population:** [ADD]

**Workflow and monitoring details**

| Monitoring | Prospective Evaluation |
|---|---|
| • | • |

**Recommendation: [OVERALL RECOMMENDATION]**
- [ADD CONTEXT FOR RECOMMENDATION]

**Next Steps**
- [ADD]

**References**
[ADD]

**Authors and Contributors (alphabetical by last name)**
[ADD]

# Monitoring

## Objective

[ADD]

## Tracking

### Model Inputs/Outputs

| Data Collection | |
| --- | --- |
| **Target Measurements** *(see Appendix)* | |

### Notes

[ADD]

### User Adherence

| Data Collection | |
| --- | --- |
| **Target Measurements** (per year after randomization / intervention) | Primary Process Measures |

### Care Delivery Outcomes

| Data Collection | |
| --- | --- |
| **Target Measurements** (per year after randomization / intervention) | Secondary Impact Measures |

## Actions

### Model Updating

| Interval | |
| --- | --- |
| **Trigger** | |
| **Decision Process** | |
| **Resolution** | |

Notes

[ADD]

## Model Retiring

| | |
|---|---|
| **Interval** | |
| **Trigger** | |
| **Decision Process** | |
| **Resolution** | |

## Execution Personnel Assignments

### Tracking Model Inputs/Outputs

- ML Ops team

### Tracking User Adherence

- [ADD]

### Tracking Care Delivery Outcomes

- [ADD]

### Model Updating

- [ADD]

### Model Retiring

- [ADD]

Stanford MEDICINE | School of Medicine

# Prospective Evaluation

## Objective

[ADD]

## Tracking

| Metrics of Success | Clinical<br><br>Financial/Operational |
|---|---|
| Study Design | |
| Interventional Study Model | |
| Masking | |
| Allocation | |
| Number of Patients | |
| Location | |
| Inclusion Criteria | |
| Exclusion Criteria | |
| Intervention | *Description and link to workflow* |
| Primary Endpoint | |
| Secondary Endpoints | **Implementation endpoints**<br><br>**Clinical endpoints** |
| Enrollment Process | |
| Assessment Schedule | |
| Duration | |
| Statistical Considerations | |

## Pre-Registration

- Recording of evaluation plan in ClinicalTrials.gov to pre-register primary outcome and study design

# Appendix

## Model Specifications

| | |
|---|---|
| **Type of task** | |
| **Outcome of interest** | |
| **Input** | |
| **Output** | |
| **Target population** | |
| **Time of prediction** | |
| **Action** | |

## Tracking Model Inputs/Outputs

| | |
|---|---|
| **Measurement** | |
| **Target** | |
| **Stratification** | |
| **Interval** | |
| **Alert Trigger** | |
| **Measurement** | |
| **Target** | |
| **Stratification** | |
| **Interval** | |
| **Alert Trigger** | |
| **Measurement** | |
| **Target** | |
| **Stratification** | |
| **Interval** | |

| Alert Trigger | |
|---|---|
| | |

## Tracking User Adherence

| Raw Data | Interval |
|---|---|
| | |
| | |
| | |
| | |

| Metric | Definition | Stratification | Interval |
|---|---|---|---|
| | | | |
| | | | |

## Tracking Care Delivery Outcomes

| Raw Data | Interval |
|---|---|
| | |

# FURM Assessment Report Example for Screening for Peripheral Arterial Disease (PAD) Use Case

Stanford MEDICINE | School of Medicine

---

### FURM Assessment Stage 1 Summary: Peripheral Arterial Disease (PAD) Prediction - Stanford Health Care

**Problem:** Many people with PAD go undiagnosed until late in their disease progression. People with undiagnosed PAD may present at the ED or to their provider with intolerance to exercise, which may require limb amputation. Furthermore, patients with PAD are at increased risk of other atherosclerotic cardiovascular events (stroke, MI), and should be placed on appropriate medical therapy (such as statins) given their elevated cardiovascular risk.

**Need:** A way to diagnose PAD earlier in a primary care population in order to increase the rate of necessary medical and surgical interventions early enough to prevent poor outcomes.

**Eligible patient population:** All patients aged 50 years or older without cancer and without a previous diagnosis of PAD, and with a primary care provider in Stanford Health Care, or possibly all such patients seen at Cardiovascular Medicine. Approximately YYYY patients meet these criteria.

---

### Proposed Model and Workflow Details

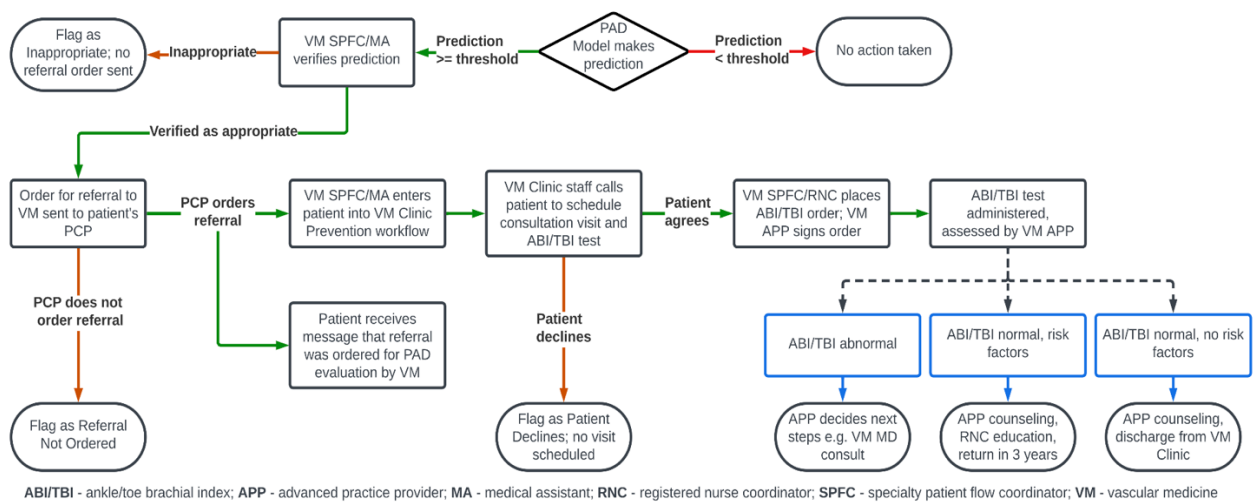

**ABI/TBI** - ankle/toe brachial index; **APP** - advanced practice provider; **MA** - medical assistant; **RNC** - registered nurse coordinator; **SPFC** - specialty patient flow coordinator; **VM** - vascular medicine

---

### Model Specification
- Outcome: PAD
- Output: probability 0-100% that a patient will develop PAD in 1 year
- Training data source: STARR OMOP
- Deployment data source: Epic Clarity

**Intervention**
- For patients with a model output above a specified probability, alert via BPA to confirm diagnosis with ABI; follow-up with their PCP and vascular specialist as needed

**Workflow Constraints**
- Personnel capacity
    - Vascular ultrasound tech: __ ABIs/day
    - Vascular surgeon/med specialist: __ patient visits/day
- Time requirements
    - __ hours for ABI
    - __ hours for MD visit

---

### Financial Projections

| Y0 (Deployment) | Y1 | Y2 | Y3 | Y4 | Y5 |
|---|---|---|---|---|---|
| $ YYYY | $ YYYY | $ YYYY | $ YYYY | $ YYYY | $ YYYY |

- Revenue: A large proportion of revenue is accrued through newly diagnosed PAD patients having cardiovascular surgery procedures.
- Costs: Model maintenance and negative margin on vascular surgery procedures comprises the majority of costs and a dynamic staffing model can lower personnel costs while maintaining capacity.

- Sensitivity: Projections are most sensitive to patient retention, model flag rate and positive predictive value.
- Of note: A significant margin contribution (~XX%) is accrued through ABI procedures on model flagged patients who are ultimately found to be false positives.

**Potential Impact for Patients or Providers**
In one month of silent deployment, the model categorized XX of YYYY screened patients (Z%) as having PAD. Given model performance metrics and sufficient personnel, this proposed workflow has potential to identify a substantial number of SHC patients with undiagnosed PAD and recommend appropriate treatment.

**Usefulness Estimates by Workflow Simulation**
In a comparison of a logistic regression, random forest and deep learning model, the deep learning model shows the strongest performance across all nursing capacity levels. However, its achievable utility caps at a nurse capacity of A patients per day when the downstream specialist's capacity is a limiting factor. Simulations also demonstrated that when doctor alert fatigue is high, the relative utility of a nurse-driven workflow is high even when nurse capacity is low.

**Ethical Considerations**
A stated goal is to decrease missed diagnoses of PAD in non-white non-male patients, but our data suggests the deployment population will be majority white. Unintended uses of disease severity or morbidity predictive tools are likely to be a recurring ethical challenge. Once introduced into the patient's chart, other actors will be interested in the prediction, and intended actors may use the prediction in unintended ways.

**Recommendation: PROCEED WITH DEPLOYMENT AND EVALUATION PLAN DESIGN**
- An appropriate technician / specialist staffing model needs to be designed to maintain profitability.
- The eligible patient population is projected to undergo decline resulting which may reduce profitability.
- It is currently unknown whether the eligible patient population will yield a non-white non-male cohort of adequate number to be powered to detect benefits/harms of the proposed workflow.

**Next Steps**
- Identify providers with capacity to execute model-guided intervention.

**References**
**PAD**

Ross EG, Shah NH, Dalman R, Nead K, Cooke J, Leeper, NJ. The use of machine learning for the identification of peripheral artery disease and future mortality risk. J Vasc Surg. 2016 Nov; 64(5): 1515–1522.

Demsas F, Joiner MM, Telma K, Flores AM, Teklu S, Ross EG. Disparities in peripheral artery disease care: A review and call for action. Semin Vasc Surg. 2022 Jun; 35(2):141–154.

Ghanzouri I, Amal S, Ho V, Safarnejad L, Cabot J, Brown-Johnson CG, Leeper N, Asch S, Shah NH, Ross EG. Performance and usability testing of an automated tool for detection of peripheral artery disease using electronic health records. Sci Rep, 2022 Aug 3; 12(1):13364.

**Evaluating AI/ML deployments**

Morse K, Bagley SC, Shah NH. Estimate the hidden deployment cost of predictive models to improve patient care. Nat Med 2020 Jan;26(1):18-19.

**Ethics for AI/ML deployments**

Char D, Shah N, Magnus D. Implementing Machine Learning in Health Care — Addressing Ethical Challenges. New Eng J Med, 2018 Mar 15;378(11):981-983.

Cagliero D, Deuitch N, Shah N, Char D. Evaluating Ethical Concerns with Machine Learning to Guide Advance Care Planning. Journal of Investigative Medicine, 2021;69:152

**Authors and Contributors (alphabetical by last name)**

Juan Banda
Alison Callahan
Danton Char
Jonathan Chen
Conor Corbin
Dev Dash
Lance Downing
Sneha Jain
Nikesh Kotecha

Jonathan Masterson
Duncan McElfresh
Keith Morse
Abby Pandya
Nigam Shah
Aditya Sharma
Rahul Thapa
Michael Wornow

# Problem, Need and Use Case Details

## Problem Statement

Many people with PAD go undiagnosed until late in their disease progression. People with undiagnosed PAD may present at the ED or to their provider with intolerance to exercise, which may require limb amputation. Furthermore, patients with PAD are at increased risk of other atherosclerotic cardiovascular events (stroke, MI), and should be placed on appropriate medical therapy (such as statins) given their elevated cardiovascular risk.

## Needs Statement

A way to diagnose PAD earlier in a primary care population in order to increase the rate of necessary medical and surgical interventions early enough to prevent poor outcomes.

## Eligible Patient Population

All patients aged 50 years or older without cancer and without a previous diagnosis of PAD (ICD9 A, B, C, D, E) and with a primary care provider in Stanford Health Care (defined by at least 1 visit within the last 18 months to a primary care clinic), or possibly all such patients seen at Cardiovascular Medicine.

## Proposed Intervention

The proposed intervention will use a machine learning classification model to estimate likelihood of PAD. For patients with a model output above a specified threshold probability, send alert via BPA to PCP to confirm diagnosis with ABI; follow-up with their PCP and vascular specialist as needed.

## Intended Impacts on Patients

- Higher rates of ABI testing
- Higher rates of ABI positivity (without smaller denominator)
- Higher rates of medical and surgical treatment for PAD
- Longer term: reduced ED visits for PAD, reduced rates of limb amputation

## Potential Unintended Impacts on Patients and Providers

- Higher rates of cardiovascular testing after PAD diagnosis
- Alert fatigue
  - In one sampled month (MM YYYY), YYYY patients were evaluated by the PAD classification model; at B% sensitivity, the model flagged XX patients as having PAD.

## Proposed Workflow

The figure below outlines the workflow for the proposed intervention. For patients with a model output above a specified probability, alert via BPA to confirm diagnosis with ABI; follow-up with their PCP and vascular specialist as needed.

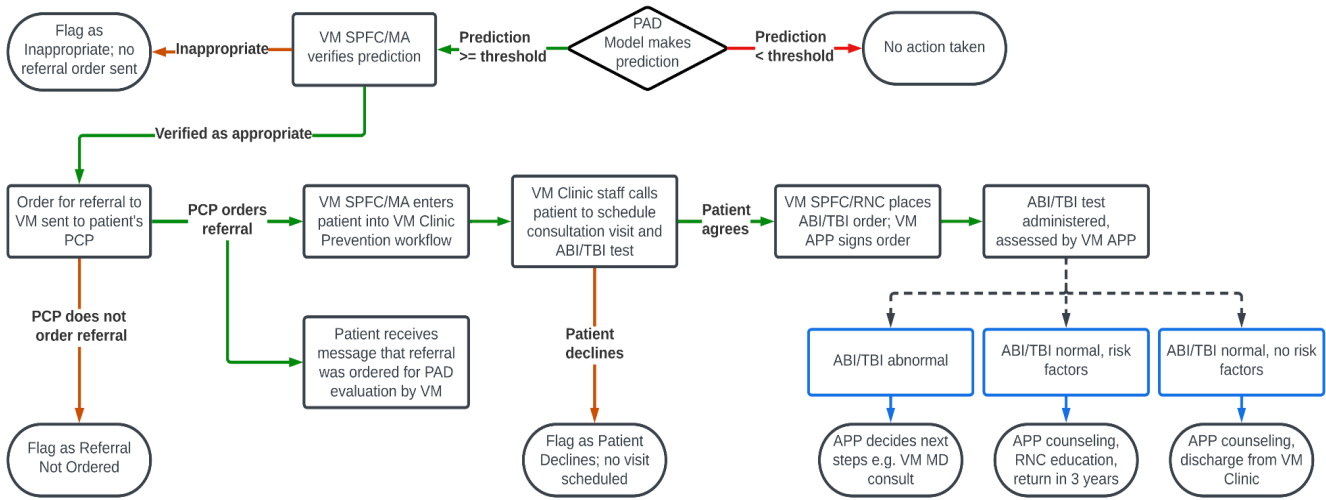

ABI/TBI - ankle/toe brachial index; **APP** - advanced practice provider; **MA** - medical assistant; **RNC** - registered nurse coordinator; **SPFC** - specialty patient flow coordinator; **VM** - vascular medicine

## Workflow Assumptions

- Capacity of personnel (e.g. calls a nurse can make per day)
  - Vascular ultrasound tech: ___ ABIs/day
  - Vascular surgeon/med specialist at some level: _ patient visits/day
- Time requirements for relevant provider activities
  - _ hours for ABI
  - _ hours for MD visit

## Utilities of Outcomes

- **False positive**: lowest cost (e.g. bleeding from ASA)
- **False negative**: low cost, may result in later diagnosis
- **True positive**: for patient - medium benefit, longer life with less disability; for health system - if management involves procedures and not only medications, there is potential for medium benefit, improving patient outcomes to avoid complex treatment later
- **True negative**: low benefit - avoids unnecessary ABI

# Usefulness assessments via simulation

## Overview

This report summarizes the results of using the APLUS tool to (1) estimate utility of a peripheral artery disease (PAD) classification model to conduct PAD screening and (2) run simulations to assess usefulness given the utility estimates and workflow properties including capacity constraints and alternative workflow parameters - specifically provider capacity in terms of number of nurses and doctors, and alert fatigue in terms of probability that a given alert is read.

## Utility Estimates

We assumed three possible outcomes for patients: "Untreated", "Medication", or "Surgery." These roughly capture the spectrum of treatment options available for patients with PAD -- either the patient's visit is concluded without treatment, the patient is prescribed medication (e.g. a statin) to reduce the risk of cardiovascular disease, or the patient undergoes a procedure like angioplasty or bypass (where we assume that patients who end up in the "Surgery" state have also been given medication prior to their procedures) [1,2]. We assume that patients who end up in the "Surgery" state have also been given medication prior to their procedures.

The utility of each of these outcomes depends on the true PAD status for a specific patient. For example, "Untreated" is the best option for patients without PAD, but has the largest cost for patients with PAD. "Medication" is the ideal outcome for patients with moderate PAD, but is undesirable for patients without PAD. "Surgery" is the most costly outcome for all patients, but the relatively best option for patients with severe PAD. We combined clinician input with utility estimates from Itoga et al. 2018 [3] to define the utilities (estimated in terms of a multiplier on remaining years living to reflect quality-adjustment on lifespan) associated with the end outcomes of each workflow in terms of a multiplier on remaining years living to reflect quality-adjustment on lifespan.

Given that a healthy patient with no PAD has a baseline utility of 1, we used the following relative utilities for various outcomes: 0.95 for patients without PAD who are prescribed medication, 0.9 for patients with PAD who are prescribed medication, 0.85 for patients with moderate PAD not prescribed medication, 0.7 for patients without PAD who undergo surgery, 0.68 for patients who have severe PAD and undergo surgery, and 0.6 for patients with severe PAD who do not undergo surgery [3].

## Usefulness Simulations

We first mapped out the states, transitions, and utilities of possible model-guided PAD screening workflows. Based on interviews with practitioners (chiefly, Elsie Gyang Ross, who is a practicing vascular surgeon), we identified two workflows to consider: (1) a nurse-driven workflow which assumes the existence of a centralized team of nurses reviewing the PAD model's predictions for all patients visiting their clinic each day; and (2) a doctor-driven workflow which assumes that the PAD model's predictions appear as a real-time alert in a patient's EHR during their visit to the clinic. These workflows are illustrated in Figure 1.

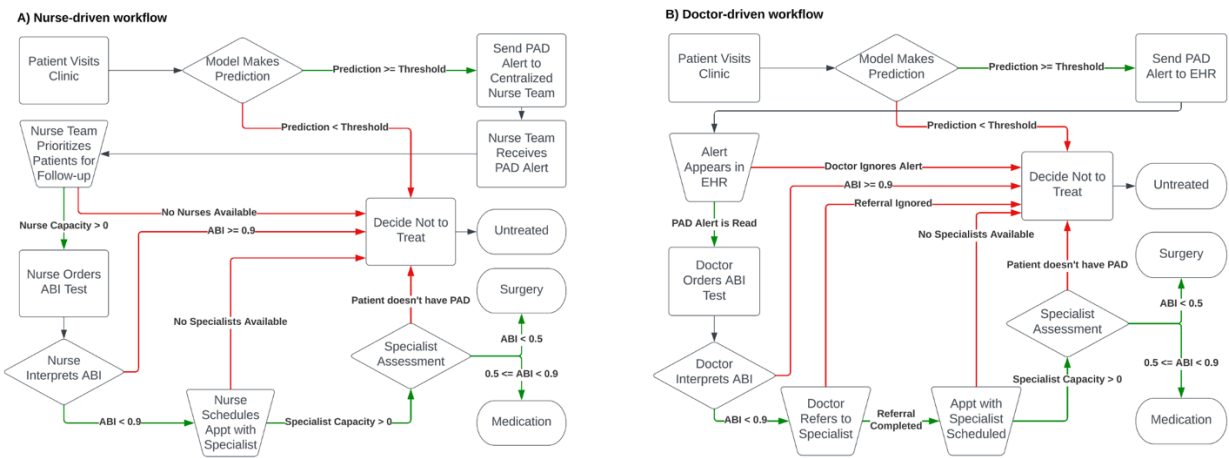

**Figure 1.** States, transitions, and transition conditions for the (a) nurse-driven workflow and (b) doctor-driven workflow. All patients begin at the "Patient Visits Clinic" state in the top left of the charts. Then, patients progress according to their individual-level properties, and end at one of 3 treatment options: "Untreated", "Medication", or "Surgery". Trapezoids represent capacity constraints, diamonds represent decision points, squares are intermediate states, and pills are end states.

In both workflows, we also assume the existence of a cardiovascular specialist who can evaluate patients after they are referred by a doctor or nurse. We assume that the specialist has a set capacity for how many patients she can see per day. However, once a patient reaches the specialist, we assume that the specialist makes the optimal treatment decision for that patient. Thus, prioritizing which patients use up the limited capacity of the specialist is the key driver of our simulated workflow's achieved utility.

The doctor-driven workflow assumes that model predictions will appear as an alert within the EHR of a patient during their visit to the clinic. If the attending physician notices this alert, she can choose to either ignore the alert or act on it. We assume that physicians ignore alerts at random. If a physician decides to act on an alert, she will either administer treatment herself (if the physician deems the patient to have a relatively normal ABI) or refer the patient to a specialist. The main constraints on this workflow are the probability that the attending physician reads the alert and the specialist's schedule.

The nurse-driven workflow assumes the existence of a centralized team of nurses tasked with reviewing the predictions of the PAD model for each patient who visits the clinic on a given day. Based on these predictions, the nursing staff decides which patients to directly refer to the specialist, thus cutting out any intermediate steps with a non-specialist physician. The main constraints on this workflow are the capacity of the nursing staff and the specialist's schedule. We assume that the nursing staff does not suffer from alert fatigue, i.e. they will not randomly ignore predictions from the PAD model. This is an assumption we have made in this study, and we acknowledge that nurses might suffer from alert fatigue as well.

We evaluated each PAD model's utility relative to three baselines: Treat None, where the model simply predicts a PAD risk score of 0 for all patients; Treat All, where the model predicts a PAD risk score of 1 for all patients; and Optimistic, where there were no workflow constraints or resource limits on model predictions. Concretely, we measured each model's expected utility achieved per patient above the Treat None baseline as a percentage of the utility achieved under the Optimistic scenario. In other words, we measured how much of the total possible utility gained from using a model was actually achieved under each workflow's constraints. The Treat None baseline should therefore always have a relative achieved utility of 0%, while all models should have a utility value of 100% in the Optimistic setting (as all patients are simply sent to the specialist for screening). For clarity, we do not show the Treat None baseline in any of the following plots, as it is always trivially set to 0%.

## Simulation of nurse-driven workflow

Specific to the nurse-driven workflow with a daily capacity of $K$, we consider two possible strategies that the nurses can leverage for processing this batch of predictions: (1) ranked screening, in which the nursing staff follows up with the K patients with the highest PAD risk scores; or (2)
thresholded screening, in which a random subset of K patients are selected from the batch of predictions whose predicted PAD risk score exceeds some cutoff threshold.

Though the deep learning model shows the strongest performance of all treatment strategies across all nursing capacity levels, its achievable utility caps out at a nurse capacity of 3 patients per day (Figure 2). This is because the downstream specialist's capacity is the limiting factor capping the achievable utility of the model. Thus, a policymaker deciding how to staff a nursing-driven workflow in which the downstream cardiovascular specialist can only see 2 patients per day could feel comfortable with staffing to a capacity of 3 patients per day, regardless of how many patients might be flagged by the model.

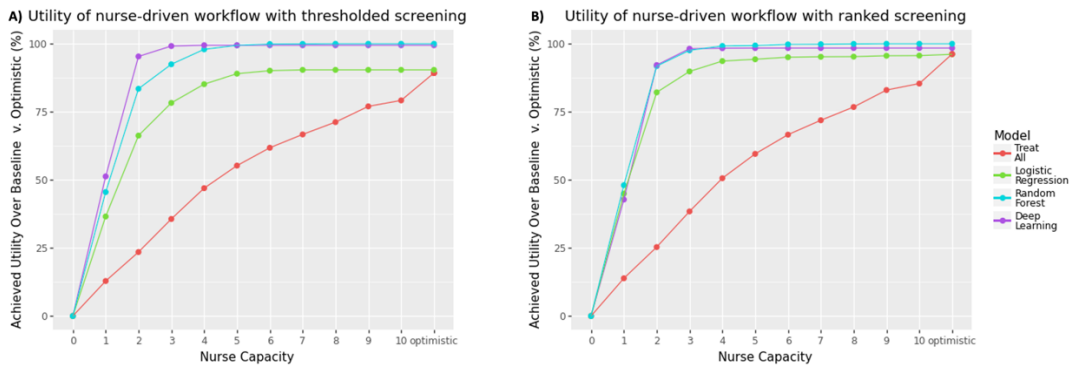

**Figure 2.** (a) Utility achieved by the nurse-driven workflow under thresholded screening across different nurse capacities using the optimal model cutoff threshold. (b) Utility achieved by the nurse-driven workflow under ranked screening across different nurse capacities. We see that the achievable utility of all models is unaffected by increases in nurse capacity beyond 3-4 nurses. All plots assume a specialist capacity of 2 patients/day.

The thresholded screening strategy (Figure 2a) shows that the deep learning model (purple line) exhibits improvements in utility over alternative treatment strategies. However, this difference becomes negligible at higher nurse capacity levels (e.g. >3 patients per day) as the limited capacity of the specialist then serves as the limiting factor on the workflow. This result is replicated under the ranked screening strategy illustrated in Figure 2b, which shows that the ML models do not exhibit substantially different achieved utility across potential nurse capacities.

## Simulation of doctor-driven workflow

In the case of the doctor-driven workflow, Figure 3 shows strong differentiation across all three ML models in the limited specialist setting. We see that the deep learning model (purple line) achieves a higher relative utility than the Treat All (red line) strategy once the probability of an alert being read is above 0.4. This is because when doctors are more likely to respond to alerts, more patients will be referred to the specialist. The specialist will then have to turn people away because of their limited capacity (set to 2 patients/day in this experiment). Thus, ensuring that we only send patients who are likely to have PAD to the specialist becomes more important as doctors become increasingly willing to act on the alerts they see (and thus exceed the capacity of the specialist to handle referrals). The accuracy of the model therefore has more influence on the workflow's utility as the probability that a doctor reads an alert increases. In our case, the deep learning model had the best predictive performance, hence the utility when using this model was greatest at higher levels of alert responsiveness.

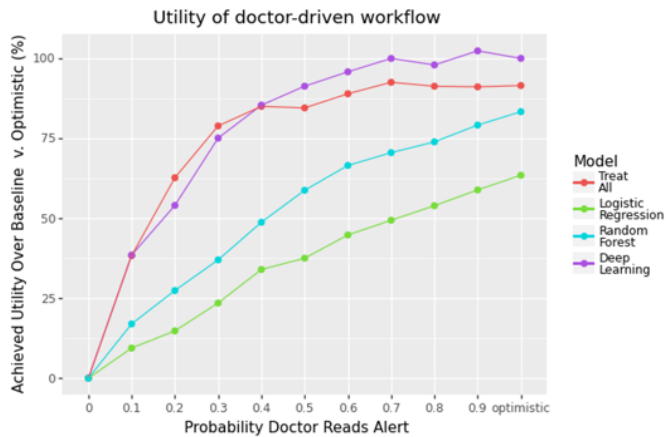

Figure 3. This plot shows the utility achieved by the doctor-driven workflow across different levels of alert fatigue using a model cutoff threshold of 0.5, assuming a specialist capacity of 2 patients/day.

## Summary

Given the results of our first two analyses, which demonstrated the superiority of the deep learning model, the final question we aimed to answer was which of the two workflows offered the optimal deployment strategy. For this experiment, we focused on quantifying the trade-off between nursing capacity and alert fatigue, as this was the primary question that came up in our conversations with clinicians. Our guiding question was as follows: How many patients would a staff of nurses need to screen per day in order to have the nurse-driven workflow yield the same expected utility as a doctor-driven workflow with a given level of alert fatigue?

To answer this question, we measured the deep learning model's achieved utility under different nurse capacities for the nurse-driven workflow using a ranked screening strategy, and compared this against the utility achieved under the doctor-driven workflow as the probability that doctors read alerts increased (where an alert was generated if the predicted probability of PAD for a patient was ≥ 0.5).

We then subtracted the latter from the former to calculate the incremental gain in achievable utility that could be expected by adopting the nurse-driven workflow at a given capacity level over a doctor-driven workflow at a given alert fatigue level. We plotted the results as a heatmap in Figure 4, under the assumption that the downstream specialist can only see 5 patients per day. The y-axis is the nursing capacity that a cell's utility value is calculated at, while the x-axis shows the level of alert fatigue in the doctor-driven workflow corresponding to that cell's measurement. Red squares (positive numbers) indicate that the nurse-driven workflow is expected to yield more utility at that capacity level than the corresponding doctor-driven workflow, while blue squares (negative numbers) indicate that the doctor driven workflow should be preferred.

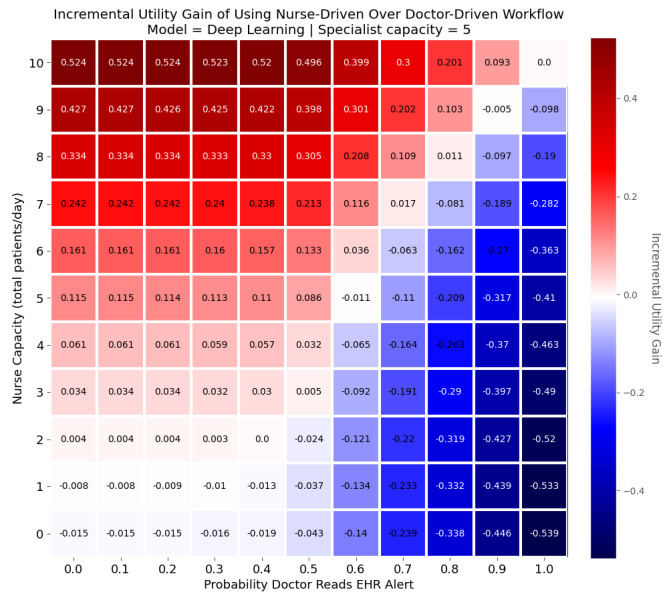

Figure 4. This heatmap shows the incremental gain from using the nurse-driven workflow over a doctor-driven workflow at a given capacity level for each workflow, assuming a specialist capacity of 5 patients/day. The y-axis represents capacity for the nurse-driven workflow, and the x-axis represents the probability that a doctor reads an EHR alert in the doctor-driven workflow. The value of the cell at coordinates (i, j) in the heatmap shows the incremental gain in achievable utility that can be expected by using a nurse-driven workflow with capacity i instead of a doctor-driven workflow with an alert fatigue level of j. Thus, positive numbers (i.e. red cells) indicate that the nurse-driven workflow is preferable to the doctor-driven workflow at their corresponding capacity levels, while negative numbers (i.e. blue cells) are situations in which the doctor-driven workflow should be preferred. That is why the top rows, which show nurse capacity at its highest, are dark red, while the far right rows, which represent the highest probability that doctors read their EHR alerts, are dark blue.

## References

[1] W.S. Aronow, Peripheral arterial disease of the lower extremities, Archives of Medical Science. 8 (2012) 375–388. https://doi.org/10.5114/aoms.2012.28568.

[2] J. Shu, G. Santulli, Update on peripheral artery disease: Epidemiology and evidence-based facts, Atherosclerosis. 275 (2018) 379–381. https://doi.org/10.1016/j.atherosclerosis.2018.05.033.

[3] N.K. Itoga, H.R. Minami, M. Chelvakumar, K. Pearson, M.M. Mell, E. Bendavid, D.K. Owens, Cost-effectiveness analysis of asymptomatic peripheral artery disease screening with the ABI test, Vascular Medicine. 23 (2018) 97–106. https://doi.org/10.1177/1358863X17745371.

Stanford MEDICINE | School of Medicine

# Financial Projections

## Overall Assessment

This assessment is preliminary, and it will be updated after we receive additional information from a pilot study. A financial projection for our initial assessment is shown below, for five years following the initial deployment year (Y0).

| Y0 (Deployment) | Y1 | Y2 | Y3 | Y4 | Y5 |
|---|---|---|---|---|---|
| $ YYYY | $ YYYY | $ YYYY | $ YYYY | $ YYYY | $ YYYY |

Recommendation - **PROCEED WITH DEPLOYMENT AND EVALUATION PLAN DESIGN**
- After Y0 (high profit), subsequent years have increasing margin due to continual outpatient volume growth
- An appropriate and technician / specialist staffing workflow needs to avoid overcapacity of providers
- False positive patients provide a large % of margin which may incur scrutiny from payors

## Assumptions

**Cohort**
- Patient cohort diagnoses / procedure count proportions over last 5 years are stable.
- Percentages of patients receiving ABI / CT / vascular surgery are stable (i.e. PAD acuity is stable).
- PAD patients are generally underdiagnosed compared to CV patients (i.e. discovery of a PAD patient leads to CV procedures, which should be counted in financial analysis). Not all newly diagnosed PAD patients will incur co-procedures but patients who are LTFU will have extensive management over the first year.

**Model & Workflow**
- Patients flagged by the model and PPV are stable at X% and X% respectively.

**Volume**
- Outpatient PCP volume YOY is projected to increase given outpatient clinic expansion at a rate of X%.
- Total addressable population projected to increase YOY over next 5 years at a rate of X%.
- Retention rate across outpatient clinics will follow stable trend (~76%).
- X% YOY capacity growth of vascular technologist / specialist.

**Financial**
- Y0 inpatient contribution margin per available bed     $YYYY
- Yearly operating cost     X%
- Yearly inflation rise     X%

## Sensitivities

| Most Sensitive to 10% increase in: | Least Sensitive to 10% increase in: |
|---|---|
| Retention Rate (X% profit drop by Y5) | Inflation |
| Model Flag Rate (X% profit increase by Y5) | Vascular Technician / Specialist Salary |
| PPV (X% profit increase by Y5) | Changes to Total Addressable Population |
| | Operating Cost % |

# Ethical Considerations

We assess the proposed intervention according to 9 ethical principles: *Responsibility*, *Equity*, *Traceability*, *Reliability*, *Governance*, *Non-Maleficence*, and *Autonomy*.

## Responsibility

*Humans should exercise judgment & remain responsible for use & outcomes.*

While apparently designed as an automated suggestion to primary clinicians, EMR research suggests automated reminders can take on unintended authority and output can be hard to challenge by end-users (i.e Cathy O'Neil demonstrating the quality of data required to rebut erroneous credit score calculations was greater than the data the scores were based on). It is unclear who will be tracking use of the PAD intervention & against what "gold standard" it will be compared.

## Equity

*Avoid unintended bias & inadvertent harm; fairness in distribution, access and benefits across patient groups.*

Though a stated goal of the PAD project is to expand the "non-white-male" data, Alison's initial data capture suggests the deployment population will be majority white. The design team for the PAD project will need to calculate whether the target population will actually yield a non-white cohort of adequate number to be powered to detect benefits/harms in those populations.

## Traceability

*Transparent & Auditable methodologies, data sources, design procedures.*

Unclear from the presentation who will be auditing or making transparent the model, as well as auditing where to best introduce the model output into the workflow & the downstream impacts from the different output introduction targets.

## Reliability

*Explicit domain of use; safety tested across the entire life cycle of use in that domain.*

The pilot domain of use appears well defined to specific primary care contexts. The pre-test probability of PAD in the target population is unclear. It has not been safety tested—or even defined what "safety testing" is appropriate for this intervention. This will likely require some projection of potential harms from introduction of other predictive tools (genome sequencing etc.) into clinical contexts. There will need to be some safeguards in place to prevent "self-fulfilling prophecies" ie. that clinical actions aren't altered in accord with the PAD prediction, before the prediction has been fully clinically validated.

## Governance

*Possess the ability to detect/avoid unintended harm & for human disengagement or deactivation.*

Unintended uses of mortality or disease severity or morbidity predictive tools are likely to be an ethical challenge. Once introduced into the patient's chart, other actors will be interested in the prediction, and intended actors may use the prediction in unintended ways.

## Non-Maleficence

*Do no harm; patient benefit; improved clinical outcomes.*

As above, a morbidity or mortality prediction may lead to an increase
in healthcare costs and an increase in ordering of interventions—particularly cardiac interventions like catheterization--that may have side effects or complications. These may lead to worse patient outcomes [1] and will need to be tracked.

## Autonomy

*Patient still in control of their healthcare; liability for AI system malfunction related to degree of autonomy; ownership of data.*

Medico-legal risks—especially liability for any adverse outcomes—are still unclear

**Quality of the Source:** Ethical assessment is based on interview with the PAD design leads. Perspectives of other stakeholders on the utility of the intervention and design choices were not captured.

**Quality of the Information:** Potential ethical challenges based on other healthcare contexts; no value collision (other than unintended uses of mortality/morbidity predictions) assessed.

## References

1.  Fenton, Joshua J., et al. "Influence of computer-aided detection on performance of screening mammography." *New England Journal of Medicine* 356.14 (2007): 1399-1409.

Stanford Medicine | School of Medicine

---

### FURM Assessment Stage 2 Summary: Peripheral Arterial Disease (PAD) Prediction - Stanford Health Care

**Problem:** Many people with PAD go undiagnosed until late in their disease progression. People with undiagnosed PAD may present at the ED or to their provider with intolerance to exercise, which may require limb amputation. Furthermore, patients with PAD are at increased risk of other atherosclerotic cardiovascular events (stroke, MI), and should be placed on appropriate medical therapy (such as statins) given their elevated cardiovascular risk.

**Need:** A way to diagnose PAD earlier in a primary care population in order to increase the rate of necessary medical and surgical interventions early enough to prevent poor outcomes.

**Eligible patient population:** All patients aged 50 years or older without cancer and without a previous diagnosis of PAD, and with a primary care provider in Stanford Health Care, or possibly all such patients seen at Cardiovascular Medicine. Approximately YYYY patients meet these criteria.

---

### Model formulation, training and testing

**Task**
- Predict a probability 0-100% that a patient will develop PAD in 1 year

**Model learning algorithm**
- Random Forest

**Data**
- Diagnosis, procedure and laboratory codes, featurized and normalized across all visits in the time window selected for each case and control.

**Testing and training splits**
- 80% train and 20% test

**Cross Validation**
- 10-fold CV

**Performance metrics**
- AUROC

**Fine-tuning procedures**
- Perform grid search for parameter optimization

**Model validation**
- Held-out test set of 20% of the data

---

### Deployment on SHC infrastructure

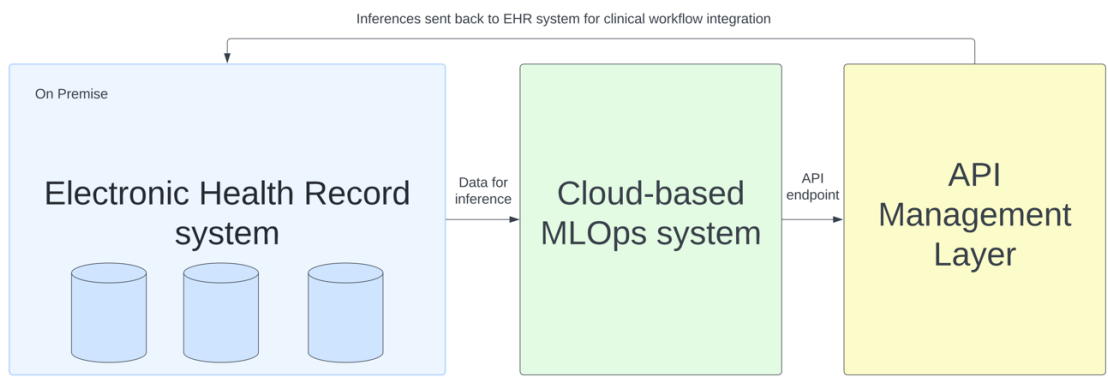

---

### Organizational integration

The workflow was developed and revised with the physician champion and service line team. The business case and resource(s) needed were approved by the service line owner. The available capacity and anticipated number of patients across sub-steps were reconciled. The effort was scoped with the TDS champion.

---

### Recommendation: PROCEED TO SILENT DEPLOYMENT
- Deployment plan has been approved by Data Science team engineering VP
- The service line owner has approved the workflow and associated resource needs

---

### Next Steps
- Design monitoring and evaluation plan

---

### References
FURM Assessment Stage 1 Summary: Peripheral Arterial Disease (PAD) Prediction - Stanford Health Care

**Authors and Contributors (alphabetical by last name)**

Juan Banda
Alison Callahan
Danton Char
Jonathan Chen
Conor Corbin
Dev Dash
Lance Downing
Sneha Jain
Nikesh Kotecha

Jonathan Masterson
Duncan McElfresh
Keith Morse
Abby Pandya
Nigam Shah
Aditya Sharma
Rahul Thapa
Michael Wornow

# Model formulation

The prediction model for PAD diagnosis follows a case/control design. There is the need for YYY cases and YYYx9 controls to build the model, to match the prevalence of PAD.

**Data labeling criteria:**

**Cohort entry criteria**: All adult patients 50 years and older with at least 1 year of EHR data.

**Cases**: Patients should have at least two separate ICD-9/ICD-10 or CPT codes for PAD and no exclusion codes. Only data collected up to 60 days prior to their diagnosis will be included in the training data to ensure codes associated with PAD diagnosis are not used in model training.

**Controls**: Patients without any diagnosis or procedure codes or text mentions for PAD in their health records.

**General exclusions:** Patients with <1 year of data.

# Model training and testing

## Model learning algorithm

Random Forest

## Data variables

The cohort described previously will use all diagnosis, procedure and laboratory codes, featurized and normalized across all visits in the time window selected for each case and control.

## Testing and training splits

80% train and 20% test

## Cross Validation

10-fold CV

## Performance metrics

AUROC

## Fine-tuning procedures

Perform grid search for parameter optimization

## Model validation

Held-out test set of 20% of the data

# Deployment on SHC infrastructure

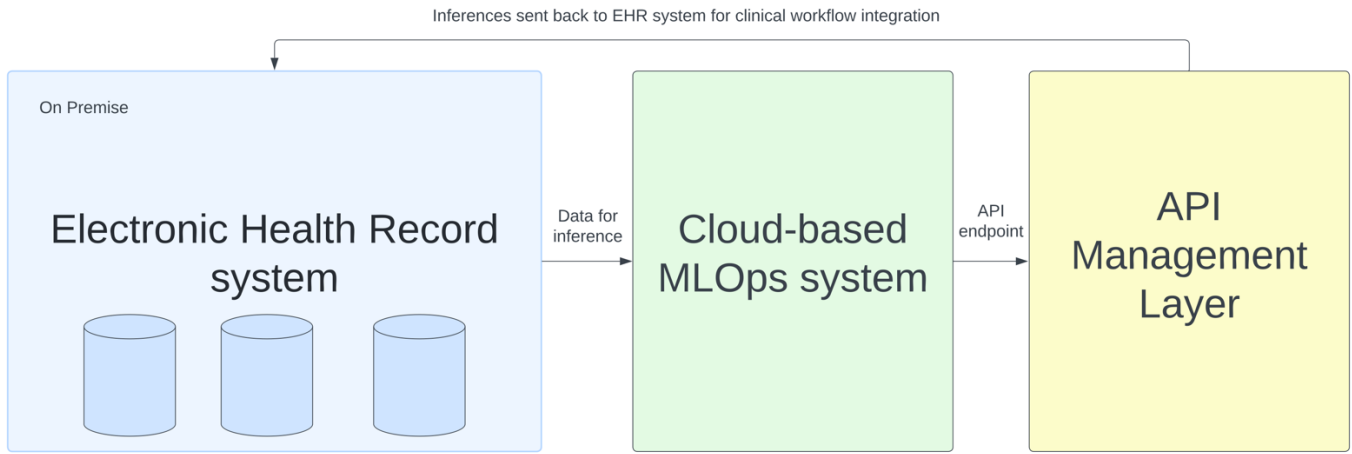

## Batch model execution

- Cloud-based MLOps system authenticates to a Managed Identity to access the on premise Electronic Health Record system and retrieve the necessary data frame.
- The model is retrieved from the MLOps model registry and executed against the data frame
  - Both the data frame and the inference are registered and saved for monitoring.
- The organization uses an API Management layer and the MLOps system feeds the inferences back into the Electronic Health Record system through this gateway.

# Organizational integration

## Clinical and operational integration

The workflow was developed and revised with the physician champion and service line team. The business case and resource(s) needed were approved by the service line owner. The available capacity and anticipated number of patients across sub-steps were reconciled. The effort was scoped with the TDS champion.

## People roles and responsibilities

| Role | Responsibilities | Name, Title |
|------|------------------|-------------|
| Service Line Owner | Budget owner who will be responsible for financial decisions | [NAME], Cardiovascular Health |
| Physician Champion | Leader and point-of-contact for the model-guided workflow | [NAME], Vascular Medicine Specialist |
| TDS Champion | Application owner responsible for IT aspects of the proposed implementation | [NAME], Director Digital Health Tech |

## Workflow

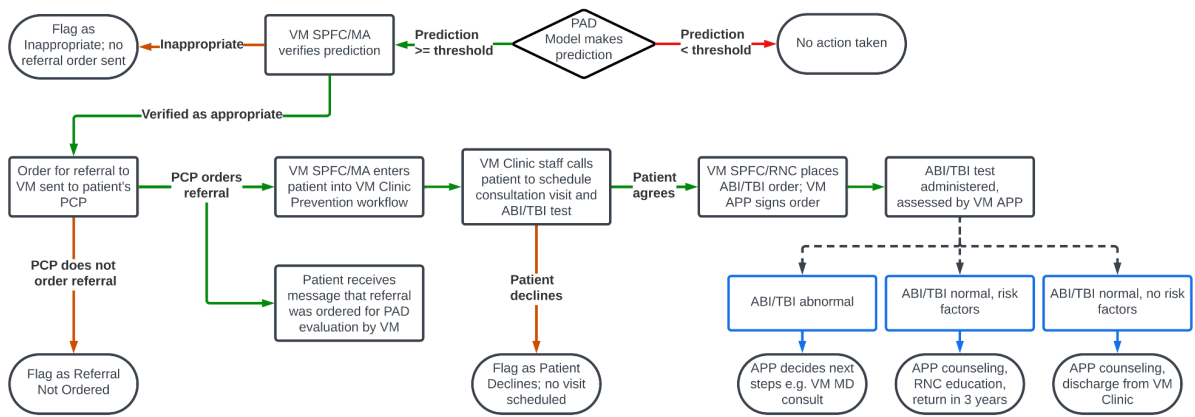

ABI/TBI - ankle/toe brachial index; APP - advanced practice provider; MA - medical assistant; RNC - registered nurse coordinator; SPFC - specialty patient flow coordinator; VM - vascular medicine

Stanford | School of Medicine
M E D I C I N E

---

**FURM Assessment Summary: Peripheral Arterial Disease (PAD) Prediction - Stanford Health Care**

---

**Problem:** Many people with PAD go undiagnosed until late in their disease progression. People with undiagnosed PAD may present at the ED or to their provider with intolerance to exercise, which may require limb amputation. Furthermore, patients with PAD are at increased risk of other atherosclerotic cardiovascular events (stroke, MI), and should be placed on appropriate medical therapy (such as statins) given their elevated cardiovascular risk.

**Need:** A way to identify undiagnosed PAD in a primary care population in order to increase the rate of necessary medical and surgical interventions early enough to prevent poor outcomes.

**Eligible patient population:** All patients aged 50 years or older without cancer and without a previous diagnosis of PAD, and with a primary care provider in Stanford Health Care, or possibly all such patients seen at Cardiovascular Medicine. Approximately YYYY patients meet these criteria.

---

**Monitoring**

- Labels become observable with significant lag (up to 1 year), thus we want to avoid reliance on these labels for monitoring.
- Sensitivity may differ as a function of time to positive label (ex: patients with diagnosed within 3-4 months of prediction may be easier to correctly predict than patients who are not diagnosed until 11-12 months after index time). We will thus track time-dependent sensitivity, stratifying patients with positive labels into discrete bins of time-to-event.
- Primary Process Measures
  - PAD Diagnosis code
- Secondary Process Measures
  - Rate of PCP agreement for patient reach-out
  - Rate of patient agreement for PAD screening
  - Rate of scheduled + completed ABI testing
  - Rate of scheduled + completed specialist visit
  - Rate of ABI abnormal results
  - Rate of medical treatment (i.e., statin prescriptions, antiplatelet prescription)
  - Number of vascular surgery treatment procedures

**Prospective Evaluation**

- Primary metric: Rate of appropriate diagnosis of PAD, and guidelines directed therapy (e.g. Statin for ASCVD).
- Study design: This is a prospective randomized quality improvement project assessing implementation and effectiveness outcomes of a machine learning assisted workflow to screen for peripheral arterial disease (PAD).
- Interventional study model: This will be a parallel study (with option to revise to step wedge) with 2 arms.
- Allocation: Randomized by Patient.
- Enrollment
- Start by predicting risk probability for all eligible patients and sort in descending order of risk
  - Randomize every other patient evaluated by MA to either intervention or control arm
  - Record control arm patients to no longer be eligible for screening for at least 18? Months
  - Deliver list of intervention arm patients to nursing workflow to queue for patient outreach
  - Repeat monthly as needed to keep nursing workflow queue full

---

**Recommendation: PROCEED WITH MONITORING AND EVALUATION**

- Monitoring and prospective evaluation plans are detailed, feasible, and with an actionable timeline

---

**Next Steps**

- Share monitoring and evaluation results with stakeholders to inform long-term planning

---

**References**

FURM Assessment Stage 1 Summary: Peripheral Artery Disease (PAD) Prediction - Stanford Health Care
FURM Assessment Stage 2 Summary: Peripheral Artery Disease (PAD) Prediction - Stanford Health Care

---

**Authors and Contributors (alphabetical by last name)**

Juan Banda
Alison Callahan
Jonathan Chen
Conor Corbin
Sneha Jain

Nikesh Kotecha
Abby Pandya
Nigam Shah
Aditya Sharma

Stanford MEDICINE | School of Medicine

# Monitoring

## Objective

Monitor the performance and user adherence of a model developed by the data science team to identify undiagnosed patients who are at risk for PAD.

## Tracking

### Model Inputs/Outputs

| Data Collection | The model will save all of its inputs / outputs to a secure server |
|---|---|
| **Target Measurements** *(see Appendix)* | <ul><li>Distribution of input features</li><li>Distribution of output probabilities</li><li>Distribution of output predicted labels</li></ul> |

#### Notes

- Labels become observable with significant lag (up to 1 year), thus we want to avoid reliance on these labels for monitoring.
  - Negative labels always take 1 year to become observable (need to wait 12 months to know a patient did not get a PAD diagnosis).
  - Positive labels may be observed sooner. Leverage this to monitor model performance metrics that condition on positive labels (i.e. sensitivity).
- Sensitivity may differ as a function of time to positive label (ex: patients with diagnosed within 3-4 months of prediction may be easier to correctly predict than patients who are not diagnosed until 11-12 months after index time).
  - We will thus track time-dependent sensitivity, stratifying patients with positive labels into discrete bins of time-to-event.
  - These metrics will be estimated continuously as labels become available.
  - Sensitivity at shorter time horizons can be estimated more frequently than sensitivity at longer time horizons. Sensitivity estimates at shorter time-horizons will be used to inform the need for model updating and retiring.
- We will track a suite of typical model discrimination and calibration measures in deployment, but we note these estimates will lag by a year and should not be relied upon when making decisions as to whether model updates or take-down is required.

### User Adherence

| Data Collection | <ul><li>Patients with predicted probability above threshold **t** will be flagged for intervention/randomization. A random subset of this list will be delivered to a workflow for review, primary care and patient outreach, and referral to vascular medicine for ABI testing and a consult.</li><li>Attempted patient outreach (i.e., intervention) will be recorded as a (Telephone, MyHealth Message, Doc Only or other) Encounter</li><li>Successful patient outreach will be reflected in an order for and completion of ABI testing and consult for the patient.</li></ul> |
|---|---|
| **Target Measurements** (per year after randomization / intervention) | Primary Process Measures<br>• PAD Diagnosis code<br>Secondary Process Measures<br>• Rate of Primary Care Provider agreement for patient reach-out<br>• Rate of patient agreement for PAD screening<br>• Rate of scheduled + completed ABI testing<br>• Rate of scheduled + completed specialist (Vascular) visit<br>• Rate of ABI abnormal results<br>• Rate of medical treatment (i.e., statin prescriptions, antiplatelet prescription)<br>• Number of vascular surgery treatment procedures |

## Care Delivery Outcomes

| | |
|---|---|
| **Data Collection** | We will estimate causal effect of the deployed model by randomizing which patients receive the outreach intervention.<br>If preliminary evidence suggests the intervention is beneficial on any measures after 3 months of parallel randomized trial design, deployment could be revised to a stepped-wedge design to retain randomization of order and causal assessment, while deliberately intending to eventually deploy to all candidate population.<br><br>All patients randomized will be logged, allowing for queries of how many have subsequent clinical procedures, diagnoses, or other significant events as outlined here. |
| **Target Measurements** (per year after randomization / intervention) | Secondary Impact Measures<br>• Number of Emergency Department encounters<br>• Number of cardiovascular surgeries (e.g., CABG)<br>• Number of cardiovascular procedures (i.e., Left heart cardiac catheterization)<br>• Number of cardiovascular tests (i.e., cardiac stress tests)<br>• MACE (Major Adverse Cardiovascular Events) (references based on ICD codes)<br>• MALE (Major Adverse Limb Events) (i.e., amputations) |

# Actions

## Model Updating

| | |
|---|---|
| **Interval** | Assessed quarterly for first year (then annually if process stabilizes) |
| **Trigger** | Deterioration in model performance / care delivery outcomes:<br>• Changes >1std in performance estimates that condition on positive labels at short time-horizons (ex: sensitivity at @3mo)<br>• Changes >1std in the distribution of input features<br>• Changes >1std in the distribution of output predictions<br>• Changes >1std in uptake or adoption rate |

| Decision Process | Applied data science team will assess whether we should (1) leave the model as is, (2) update the model, or (3) recommend model retirement. |
|---|---|
| Resolution | ML Ops team will inspect the data pipeline feeding into the model. Any changes in ETLs, feature definitions, or prediction trigger events occurring between the silent deployment period and production will be noted.<br><br>• If no changes are discovered, the data science team will speak with the clinical users of the model to discuss changes in model outputs to detect data drift.<br><br>If the distribution of patients is changing (e.g. we are seeing sicker patients) while the model itself is acting properly (e.g. sensitivity @3mo which should remain stable under label shift\), the decision should be to leave the model as is.<br><br>• Alternatively, the model could be retrained with more recent examples<br>• However, due to the 1 year lag in observable labels, re-training a newly deployed model will be difficult within reasonable follow-up time frames<br>• Future updated models may instead use ABI test results as a reference target while accounting for censoring of untested cases. |

Notes

• Labels become observable with a large lag (maximum 1 year)
• Cannot rely on common model performance criteria (e.g. AUROC, AUPRC, calibration) or a continuous usefulness assessment (which requires an estimate of true positives, false positives, true negative, and false negatives) to decide if a model needs updating.

## Model Retiring

| Interval | Annually |
|---|---|
| Trigger | Model updating discussion concluded that the model is neither retrainable nor performant enough to justify its continued use |
| Decision Process | Data science team, provider champions, and service line owner will meet to decide if model should be retired |
| Resolution | Model will be archived, (randomized) patient lists delivered to workflow will cease. |

## Execution Assignments

### Tracking Model Inputs/Outputs

• ML Ops team

### Tracking User Adherence

• Data Science and *PROVIDER CHAMPION*

### Tracking Care Delivery Outcomes

• Data Science and *PROVIDER CHAMPION*

## Model Updating

- Data Science and ***PROVIDER CHAMPION***

## Model Retiring

- Data Science, ***PROVIDER CHAMPION and SERVICE LINE OWNER***

# Prospective Evaluation

## Objective

To assess the implementation and effectiveness of a machine learning-assisted workflow to screen for PAD.

## Tracking

| | |
|---|---|
| **Metrics of Success** | Increased appropriate diagnosis of PAD, and guidelines directed therapy (e.g. Statin for ASCVD). |
| **Study Design** | This is a prospective randomized quality improvement project assessing implementation and effectiveness outcomes of a machine learning assisted workflow to screen for PAD. |
| **Interventional Study Model** | This will be a parallel study (with option to revise to step wedge) with 2 arms. |
| **Masking** | Not Applicable/Unblinded |
| **Allocation** | Randomized by Patient |
| **Number of Patients** | TBD |
| **Location** | Stanford Health Care System |
| **Inclusion Criteria** | <ul><li>Age > 50</li><li>Has visit at Stanford Health Center in last 24 months</li></ul> |
| **Exclusion Criteria** | <ul><li>Metastatic Cancer diagnosis</li><li>Previous PAD Diagnosis code</li><li>ABI Test done within past 18 months</li><li>Previous Vascular Surgery visit within past 18 months</li><li>Inference and intervention / randomization of patient within past 24 months (avoid repeated contacts)</li></ul> |
| **Intervention** | See Description and workflow in FURM Assessment Stage 2 Summary: Peripheral Arterial Disease (PAD) Prediction - Stanford Health Care |
| **Primary Endpoint** | PAD Diagnosis |
| **Secondary Endpoints** | <ul><li>Implementation endpoints<ul><li>Not Participant - Sanity Criteria Not Met</li></ul></li></ul> |
|  | - o Not Participant - PCP Confirmation Not Met<br>  o Not Participant - Patient Agreement Not Met<br>  o Participant - Patient Agrees<br>  o Participant Results: ABI Normal No Health Risk Factors<br>  o Participant Results: ABI Normal Positive Health Risk Factors<br>  o Participant Results: ABI Abnormal<br>• Starting statin in statin-naive patients<br>• Prescription rates of other medicines<br>  o Antiplatelets—per PAD guidelines, not much evidence of asymptomatic PAD<br>  o Antihypertensive therapies<br>• Surgical Treatment<br>• Biological Cardiovascular Risk Factors<br>  o Lipid panel collected within past 2 years<br>  o LDL < 70 at latest check<br>  o HbA1c checked within past 2 years<br>• Healthcare Resource Utilization in past 12 months<br>  o Cardiology visits<br>  o PCP visits<br>  o Resting echocardiograms<br>  o Stress tests or CCTA<br>  o Coronary angiograms<br>  o CABG<br>  o Carotid ultrasound<br>• Quality of life<br>• All-cause death<br>• Longer term outcomes (for future considerations beyond 12-month timeframe)<br>  o Major Adverse Cardiovascular Events<br>    ▪ Myocardial infarction<br>    ▪ Stroke<br>    ▪ Cardiovascular Death<br>  o Major Adverse Limb Events<br>    ▪ Severe Limb Ischemia leading to intervention<br>    ▪ Major Vascular Amputation |
| **Enrollment Process** | • Start by predicting risk probability for all eligible patients and sort in descending order of risk<br>• Randomize every other patient evaluated by MA to either intervention or control arm<br>• Record control arm patients to no longer be eligible for screening for at least 18? Months<br>• Deliver list of intervention arm patients to nursing workflow to queue for patient outreach<br>• Repeat monthly as needed to keep nursing workflow queue full |
| **Assessment Schedule** | TBD |
| **Duration** | 12 months |
| **Statistical Considerations** | Power Calculation: TBD |

Stanford | MEDICINE | School of Medicine

# Appendix

## Tracking Model Inputs/Outputs

| Measurement | Distribution of input features |
|---|---|
| Target | Distribution of demographics, counts of diagnoses, procedures, medication prescriptions, and lab orders within the target population |
| Stratification | None |
| Interval | Monthly |
| Alert Trigger | Deviations of more than 1 standard deviation in the distribution of a feature compared to the Retrospective Test Set |

| Measurement | Distribution of output probabilities |
|---|---|
| Target | Distribution of raw probabilistic outputs of the Model |
| Stratification | Demographic group (age, race, sex, ethnicity) |
| Interval | Month |
| Alert Trigger | >1 standard deviation change in mean |

| Measurement | Distribution of output predicted labels |
|---|---|
| Target | Distribution of thresholded predictions output by the Model |
| Stratification | Demographic group (age, race, sex, ethnicity) |
| Interval | Monthly |
| Alert Trigger | >1 standard deviation change in mean |

## Tracking User Adherence

| Raw Data | Interval |
|---|---|
| Number of Patient Outreach in Intervention Population | Monthly |
| Number of Orders (Referral to Vascular Medicine and ABI test) in Patients Receiving Intervention | Monthly |
| Number of ABI Tests Completed in Patients Receiving ABI Order | Monthly |
| Number of Patients with ABI Abnormal | Monthly |

| Metric | Definition | Stratification | Interval |
|---|---|---|---|
| Uptake | Percentage of Higher Risk Patients Randomized to Intervention Arm receiving an Outreach + ABI Order | | Monthly |
| Adoption rate | Percentage of Intervention patients receiving an outreach who complete an ABI test | | Monthly |

Tracking Care Delivery Outcomes

| Raw Data | Interval |
|---|---|
| Percentage of patients with positive ABI test receiving medication (i.e., statin) and vascular surgery | Monthly |